\documentclass[paper,nofootinbib] {revtex4}

\usepackage{graphicx}
\usepackage{slashbox}
\usepackage{epsfig}
\usepackage{fancyhdr}
\usepackage{mathbbol}
\usepackage{pstricks}
\usepackage{color}
\usepackage{bbm}
\usepackage{multirow}
\usepackage{axodraw}
\usepackage{bm}
\usepackage{amsmath}
\usepackage{amsfonts}
\renewcommand{\arraystretch}{1.5}

\newcommand{\field}[1]{\mathbb{#1}}

\def\be{\begin{equation}}
\def\ee{\end{equation}}
\def\bea{\begin{eqnarray}}
\def\eea{\end{eqnarray}}
\def\M{{\cal M}}

\begin{document}

\title{Strong Couplings of $X(3872)_{J=1,2}$ \\ and a New Look at $J/\psi$ Suppression in Heavy Ion Collisions}


\author{F Brazzi$^{\ddag}$, B Grinstein$^{\diamond}$, F Piccinini$^{\dag}$, AD Polosa$^{\square,\P}$ and C Sabelli$^{\ddag,\P}$} 
\affiliation{
$^\ddag$Dipartimento di Fisica, Universit\`a di Roma `Sapienza', Piazzale A. Moro 2, Roma, I-00185, Italy\\
$^{\diamond}$University of California, San Diego, 9500 Gilman Drive, La Jolla CA 92093, USA\\
$^\dag$INFN Pavia, Via A. Bassi 6, Pavia, I-27100, Italy\\
$^\square$Dipartimento di Fisica e di Medicina Molecolare, Universit\`a di Roma `Sapienza', Piazzale A. Moro 2, I-00185, Italy\\
$^\P$INFN Roma, Piazzale A. Moro 2, Roma, I-00185, Italy}

\begin{abstract}
We define and compute from data the strong couplings of the $X(3872)$ with both of the possible quantum
numbers assignments $J^{PC}=1^{++},2^{-+}$.  
We use these to compute cross sections for $J/\psi$ resonance scattering into $D\bar{D}^*$. 
As an application of the results obtained we revise the calculation of the $J/\psi$ absorption in a hot hadron gas to confront with recent RHIC observations in Au-Au collisions.
\end{abstract}


\maketitle
%\tableofcontents

\thispagestyle{fancy}

\section{Introduction}
Recently the BaBar collaboration took issue with the statement that the $X(3872)$ is a $1^{++}$ resonance, as was widely accepted,  raising the hypothesis of $2^{-+}$ quantum 
numbers~\cite{babarbn}, yet to be confirmed by Belle or LHC experiments.

In consideration of this we have discussed in a previous paper~\cite{burns} the consequences of  the quantum numbers assignment suggested by BaBar confronting the expected mass of the $2^{-+}$ standard charmonium with the one measured for the $X$. We also raised the problem of the small $X$ prompt production cross-section predicted for a 
$1^1D_2$ charmonium when confronted with the measured one. Indeed, assuming the $2^{-+}$ hypothesis, the molecular interpretation of the $X(3872)$ (at odds with data also in the $1^{++}$ case, as discussed in~\cite{Bignamini,Bignamini2} and contested in~\cite{Artoisenet}) is ruled out and the charmonium interpretation comes back into play. Also the tetraquark is still an open option~\cite{burns}.

In this paper we will expand on the consequences of the $J^{PC}$ assignment studying the decay modes of the $X(3872)$ under the hypothesis that it is a $1^{++}$ state, 
let us call it $X_1$, or a $2^{-+}$ state, call it $X_2$, but making no hypotheses on its structure (charmonium, molecule, tetraquark). 

As a first step, we will define a general parameterization of the transition matrix elements, describing the known decays  in terms of a set of strong coupling constants. Using the experimental information available and summarized in Table~\ref{tab:expdata}, as discussed in~\cite{SpecNC}, we will determine the strong couplings in the $1^{++}$ and $2^{-+}$ cases by the explicit computation of decay widths. In our fits we will also use the data on $X\to J/\psi\; \omega$ decays reported in~\cite{babarbn}.

\begin{table}[!h]
  \begin{tabular}{||l|c|c||}
    \hline
 Decay Mode   & $\mathcal{B}$ & $\sigma(\mathcal{B})$\\
    \hline
    $X\to J/\psi\:\pi^+\pi^-$ &$0.055$  & $0.020$\\
    \hline
    $X\to J/\psi\;\pi^+\pi^-\pi^0$ &$0.045$  & $0.030$\\
    \hline
    $X\to J/\psi\;\gamma$ &$0.0135$  & $0.0060$\\
    \hline
    $X\to D^0\bar{D}^{0*}$ &$0.67$  & $0.13$\\
      \hline
  \end{tabular}
\label{tab:expdata}
\caption{Branching ratios $\mathcal{B}$ and one sigma errors $\sigma(\mathcal{B})$ for the observed decays of $X(3872)$~\cite{SpecNC}.}
\end{table}

We will not attempt any theoretical determination of the strong coupling constants we are going to define. This would require to formulate some hypotheses on the structure of the $X$ and the use of approaches such as quark models or QCD sum rules. This work could be done elsewhere and confronted with the coupling strengths found here.

We also confirm that using data in~\cite{babarbn}, the negative parity assignment for the $X$ is indeed favored: as opposite to an earlier analysis by CDF~\cite{cdfangular} on the $J/\psi\;\pi^+\pi^-$ angular distribution, indicating that both the $1^{++}$ and $2^{-+}$ assignments are equally possible, we will show that the $2^{-+}$ assignment would be the preferred one. 

Once we have extracted from data the strong couplings of the $X_{1,2}$ to the $\omega$ and $\rho$ vector mesons, 
we will calculate the $J/\psi \; (\rho,\omega)\to X_{1,2}\to D^0\bar D^{0*}$ cross-section.
This calculation could be of relevance to the study of a classic background to the $J/\psi$ suppression signal in heavy ion collisions: the $J/\psi$ can be absorbed by a hot gas of  hadrons including pions and the lighter vector resonances at a similar rate as it is supposed to occur in a phase of deconfined quarks and gluons. If one finds that this mechanism is effective at temperatures smaller than the Hagedorn temperature, considered as the limiting one for the hadronic 
matter~(see {\it e.g.}~\cite{cabibboparisi}), then the $J/\psi$ suppression signal should be regarded as the less compelling among the many indications available of a new phase of matter produced in heavy ion collisions.

We will show that only the $X_2$ would contribute effectively to the $J/\psi \;(\rho,\omega)\to D^0\bar D^{0*}$ absorption, whereas the $X_1$ has no significant contribution.
It is clear that given the narrowness of the $X(3872)$, the process of converting $J/\psi\; (\rho,\omega)$ into $D^0\bar D^{0*}$ due to an intermediate $X$ is too slow to have a mean free path 
in a gas of $\rho$'s and $\omega$'s that is likely larger than the typical size expected for the fireball generated in a heavy ion collision. 

Most of the studies on the hadronic suppression of $J/\psi$ are made under the hypothesis that the in-medium $D$ and $D^*$ mesons have a reduced mass and a larger width~\cite{Faessler:2002qb, Fuchs:2004fh,Tolos:2007qr,Tolos:2007vh,Gottfried:1992bz,Hayashigaki:2000es,Kumar:2009xc,Kumar:2010gb,Kumar:2010nz}. This in turn would compensate for the narrow width of the $X$, because of a larger $D^0\bar D^{0*}$ phase space -- we find that mass variation of the open charm mesons allows a sizable contribution of the $X_2$ in the hadronic $J/\psi$ suppression. It has been claimed that if one sets a definite temperature at which the $D$ and $D^*$ masses drop down, one could reproduce the dip observed in $J/\psi$ production data in correspondence of a critical centrality of the nucleus-nucleus collision~\cite{Burau:2000pn,Blaschke:2002ww}. As discussed in the section on the comparison to data on $J/\psi$  suppression at RHIC, we do not confirm this pattern in our analysis.

We will add the contribution of the $X_2$ to former results on the non-resonant hadronic $J/\psi$ suppression due to a hot $\pi/\eta,\rho/\omega,\phi,K^{^{(*)}}$ gas and show its relative weight with respect to that. 
This study is intended to be  suggestive of the fact that the $XYZ$ resonances could be of relevance as intermediate states in a number of physical processes, as proposed for example in~\cite{Bigi}.

In Section~\ref{sec:decays} we define the strong coupling constants for $X_1$ and $X_2$ and calculate their values using available data on the $X(3872)$ decay modes.
In Section~\ref{subsec:xsect}, using the results obtained in Section~\ref{sec:decays}, we compute the cross-sections for processes like $J/\psi (\rho,\omega)\to X_{1,2}\to D^0\bar D^{0*}$. 
Section~\ref{subsec:data} is devoted to the computation of the average absorption length of the $J/\psi$ in a hadron fireball, taking into account the in-medium properties of open charm mesons described in Section~\ref{subsec:medium}. We compare these results with those obtained in absence of effects modifying the mass and width of $D$ mesons. We include in our analysis resonant and non-resonant contributions neglecting their interference. Finally in Section~\ref{subsec:hagedorn} we compare our predictions to the RHIC data
on Au-Au collisions.

\section{$X$ decays}
\label{sec:decays}
We start with the parameterization of the transition matrix elements for the decay processes in Table~\ref{tab:expdata} in terms of coupling strengths whose numerical values are then extracted by comparison with experimental data.
In the next subsection we discuss the $M_{fi}$ matrix elements which are related to the transition matrix elements $T_{fi}$ by
\be
T_{fi}=(2\pi)^4\delta^{4}(p_i-\sum_f p_f )\; M_{fi}.
\ee

As for the normalization of states in $M_{fi}$ the standard $1/\sqrt{2EV}$ is used.

\subsection{Transition matrix elements}
We require that strong transition matrix elements are parity even Lorentz scalars obtained by combining the momenta and polarizations of the initial and final particles. The conservation of angular momentum fixes the decay wave of $A\to BC$: $J_A=(J_B\oplus J_C)\oplus \ell_{BC}$,
$\ell_{BC}$ being the relative orbital angular momentum in the final state. For each unit of orbital angular momentum in the final state
there must be factor of a spatial component of the momentum in the transition matrix element. 
Here and in the following we use the notation $\psi$ and $J/\psi$ interchangeably.

{\bf The $\bm{J^{PC}=1^{++}}$ case}. The decay $X\to \psi V$, with $V=\rho,\omega$ is an $\ell=0$ decay, since from the point of view of the $J^P$ quantum numbers it corresponds to $1^{+}\to 1^{-}1^{-}$. 
There is only one combination of momenta and polarizations which has all the properties we enumerated 
above~\footnote{In the rest frame of the decaying particle $P_\sigma=(m_{_X},\bm{0})$ and one can write
\begin{displaymath}
\langle \psi(\epsilon,p) V(\eta,q)|X(\lambda,P)\rangle = g_{_{1\psi V}}\;m_{_X}\;\epsilon^{ijk0}\;\lambda_i(P)\;\epsilon^*_j(p)\;\eta^*_k(q) = g_{_{1\psi V}}\;m_{_X}\; \left(\bm{\lambda}(P) \times \bm {\epsilon}^*(p)\right)\,\cdot \bm{\eta}^*(q) 
\end{displaymath}
which is the scalar product of two polar vectors, the first coming from the vector product between an axial vector $\bm{\lambda}$ and a polar vector $\bm{\epsilon}$. Moreover the above expression does not contain any spatial component of the momenta and thus accounts for an {\it S}~-~wave process.}
\be
\langle \psi(\epsilon,p) V(\eta,q)|X(\lambda,P)\rangle = g_{_{1\psi V}}\;\epsilon^{\mu\nu\rho\sigma}\;\lambda_\mu(P)\;\epsilon^*_\nu(p)\;\eta^*_\rho(q)\;P_\sigma.
\ee

The decay $X\to D^0\bar{D}^{0*}$ is also an $\ell=0$ process, since it corresponds to $1^{+}\to 0^{-}1^{-}$.
The matrix element  can be written in terms of a second coupling strength, $g_{_{1DD*}}$, as follows
\be
\langle D^0(p) \bar{D}^{0*}(\epsilon,q)|X(\lambda,P)\rangle= g_{_{1DD^*}}\;\lambda^\mu(P)\;\epsilon^*_\mu(q).
\ee
In order to conserve charge conjugation one should consider the final state $D^0\bar{D}^{0*}+\bar{D}^0D^{0*}$. 
As explained in Sec.~\ref{app:multiplicityres} of the Appendix, we can consider only the $D^0\bar{D}^{0*}$ component of the final state in what follows.

{\bf The $\bm{J^{PC}=2^{-+}}$ case}. In this case, both the decays $X\to \psi V$ and $X\to D^0\bar{D}^{0*}$ are $\ell=1$ processes, since they correspond to $2^{-}\to 1^{-}1^{-}$ and $2^{-}\to 0^{-}1^{-}$ transitions respectively.

The spin of the $X$ is described by a symmetric traceless polarization tensor $\pi^{\mu\nu}$ satisfying  $P_\mu\pi^{\mu\nu}=0$.
In the rest frame the five independent components can be set in a $3\times 3$ traceless tensor $\pi^{ij}$. 
For the sum over polarizations we have \cite{Feynman:1996kb}
\bea
 \sum_{\rm pol}\pi_{\mu\nu}(k)\pi^*_{\alpha\beta}(k)&=& \frac{1}{2}\left( g_{\mu\alpha}g_{\nu\beta}
 +g_{\mu\beta}g_{\nu\alpha}
  -g_{\mu\nu}g_{\alpha\beta}\right) 
\label{propg}
  -\frac{1}{2m^2}\left( 
g_{\mu\alpha}{k_\nu k_\beta}
 +g_{\nu\beta}{k_\mu k_\alpha}+
  g_{\mu\beta}{k_\nu k_\alpha} +g_{\nu\alpha}
  {k_\mu k_\beta}\right) \\ \nonumber
 &+& \frac{1}{6}
( g_{\mu\nu} 
    +\frac{2}{m^2}    k_\mu k_\nu  )
( g_{\alpha\beta} 
    +\frac{2}{m^2}    k_\alpha k_\beta  ),
\eea
with $k^2=m^2$.

For the decay $X\to\psi V$, we have to determine the transition matrix element $\langle \psi (\epsilon,p) V(\eta,q)|X(\pi,P)\rangle$. There are two ways of combining momenta and polarizations which give a parity even Lorentz scalar\footnote{An $\epsilon^{\mu\nu\rho\sigma}$ tensor is needed to obtain even parity. Moreover one cannot contract the two indices of the symmetric $\pi$ tensor with two of the indices of the completely antisymmetric $\epsilon$ tensor.}:
\begin{itemize}
\item[({\it i})] a polarization vector contracts with the left index of the $\pi$ tensor. If that of the $\psi$ we have
\be
\label{uno}
\epsilon^{*\alpha}(p)\:\pi_{\alpha\mu}(P)\;\epsilon^{\mu\nu\rho\sigma}\;p_\nu\;q_\rho\;\eta^*_\sigma(q)
\ee
or if the $V$ one does
\be
\label{due}
\eta^{*\alpha}(p)\:\pi_{\alpha\mu}(P)\;\epsilon^{\mu\nu\rho\sigma}\;q_\nu\;p_\rho\;\epsilon^*_\sigma(q);
\ee
\item[({\it ii})] a momentum contracts with the left index of the $\pi$ tensor. One can have
\be
\label{tre}
p^\alpha\:\pi_{\alpha\mu}(P)\;\epsilon^{\mu\nu\rho\sigma}\;q_\nu\;\epsilon^*_\rho(p)\;\eta^*_\sigma(q)
\ee
\noindent  and the remaining combinations of momenta obtained by replacing $pq$ by $pp$, $qq$, $qp$.
\end{itemize}
Since we have a {\it P}~-~wave decay, we should not have non-zero terms proportional to $p_i q_j$, where $i$ and $j$ are spatial indices.
In the $X$ rest frame such terms are absent in Eq.~(\ref{uno}) and~(\ref{due})  since they would be proportional to ${\bf p}\times {\bf q}$ -
which vanishes only in the rest frame of the decaying particle.
The only non zero combination of the type of Eq.~(\ref{tre}), not containing the $p_i q_j$ terms is
\be
Q^\alpha\:\pi_{\alpha\mu}(P)\;\epsilon^{\mu\nu\rho\sigma}\;P_\nu\;\epsilon^*_\rho(p)\;\eta^*_\sigma(q),
\ee
where $Q=p-q$ and $P=p+q$. 

In conclusion we find that there are only three invariant amplitudes one can form by combining these tensors
\be
T_1=\epsilon^{*\alpha}(p)\:\pi_{\alpha\mu}(P)\;\epsilon^{\mu\nu\rho\sigma}\;p_\nu\;q_\rho\;\eta^*_\sigma(q)+\eta^{*\alpha}(q)\:\pi_{\alpha\mu}(P)\;\epsilon^{\mu\nu\rho\sigma}\;p_\nu\;q_\rho\;\epsilon^*_\sigma(p),
\ee
\be
T_2=\epsilon^{*\alpha}(p)\:\pi_{\alpha\mu}(P)\;\epsilon^{\mu\nu\rho\sigma}\;p_\nu\;q_\rho\;\eta^*_\sigma(q)-\eta^{*\alpha}(q)\:\pi_\alpha\mu(P)\;\epsilon^{\mu\nu\rho\sigma}\;q_\nu\;p_\rho\;\epsilon^*_\sigma(p),
\ee
\be
T_3=Q^\alpha\:\pi_{\alpha\mu}(P)\;\epsilon^{\mu\nu\rho\sigma}\;P_\nu\;\epsilon^*_\rho(p)\;\eta^*_\sigma(q).
\ee
\noindent 
which carry three implicit polarization indices.
The first two correspond to the sum and the difference of Eq.~(\ref{uno}) and~(\ref{due}), which it turns out to be useful to further reduce the number of independent tensors. 
Indeed one can show that $T_1$ and $T_3$ are one and the same tensor. To do this  we prove that the following relation among sums over polarizations holds
\be
\label{magicrelation}
\big(\sum_{{\rm pol}}T_1T_3^*\big)^2=\big(\sum_{{\rm pol}}|T_1|^2\big)\big(\sum_{{\rm pol}}|T_3|^2\big).
\ee
The above condition implies that the two tensors are equal up to a constant if the sum over polarizations has the properties of an inner product.
The Schwarz inequality states indeed that for all vectors $\bm{v}$, $\bm{w}$ 
\be
|\langle \bm{v},\bm{w}\rangle|^2\leq\langle \bm{v},\bm{v}\rangle\langle \bm{w},\bm{w}\rangle,
\ee
\noindent where $\langle \cdot,\cdot\rangle$ is an inner product: the equality holds only if the two vectors are linearly dependent, {\it i.e.}, if they are parallel.
Given two vectors $\bm{v}$, $\bm{w}\in\field{C}^n$, the inner product is defined as $\langle\bm{v},\bm{w}\rangle=\sum_{\substack{n=1}}^N v_nw_n^*=\sum_{\substack{n=1}}^N v_n^*w_n$.
Here we are evaluating sums over polarizations, labeled by $n$, which means
\be
\sum_{{\rm pol}}T_iT_j^*=\sum_{\substack{n=1}}^{5\times3\times3}T_i^{(n)}(T_j^{(n)})^*,
\ee
\noindent where we are summing over the $5$ polarizations of the $X$ and the $3$ of the vectors.
Therefore Eq.~(\ref{magicrelation}) implies that for each polarization configuration, $T_1$ and $T_3$ are equal up to a constant,
and we can choose one out of the two for our basis of linearly independent tensors. We choose to keep $T_3$ and eliminate $T_1$. 
The final choice for the parameterization is
\be
\langle \psi(\epsilon,p) V(\eta,q)|X(\pi,P)\rangle = g_{_{2\psi V}}T_2 + g_{_{2\psi V}}^\prime T_3,
\ee
where $V=\rho,\omega$.

Finally we consider $X\to D^0\bar{D}^{0*}$. One can easily build a parity even Lorentz  scalar by contracting the $\pi$ tensor 
with the $D^{0*}$ polarization vector and the $D^0$ momentum\footnote{
The even parity can be easily understood. In the rest frame of the $X$ one has
\begin{displaymath}
\langle D^0(p) \bar{D}^{0*}(\epsilon,q)|X(\pi,P)\rangle= g_{_{2DD^*}}\pi^{ij}\epsilon^*_i(q) \,p_j=g_{_{2DD^*}} \left(\bm{a}\cdot\bm{\epsilon}^*(q)\right)\otimes \left(\bm{v}\cdot\bm{p}\right)
\end{displaymath}
where $\bm{a}$ and $\bm{v}$ are an axial and a polar vector respectively defined by $\pi^{ij}=a^{i}\otimes v^j+v^{i}\otimes a^j$.
}
\be
\langle D^0(p) \bar{D}^{0*}(\epsilon,q)|X(\pi,P)\rangle= g_{_{2DD^*}}\pi^{\mu\nu}\epsilon^*_\mu(q) p_\nu.
\ee

\subsection{Decay widths and determination of the strong couplings} 
{\bf The $\bm{J^{PC}=1^{++}}$ case}. Since $\omega$ and $\rho$ have different isospin quantum numbers
in principle one needs to use different couplings to describe these decays: 
$g_{_{1\psi\omega}}$ and $g_{_{1\psi\rho}}$.
To determine these two values we write the partial decay widths for $X\to J/\psi\;\rho$ and $X\to J/\psi\;\omega$ as in Eq.~(\ref{gammacompletarho}) and (\ref{gammacompletaomega}) in Appendix~\ref{app:rho} and \ref{app:omega} respectively. For $X\to J/\psi\;\rho\to J/\psi\; \pi^+\pi^-$ we have
\be
\Gamma(X\to J/\psi\;\pi^+\pi^-)=\frac{1}{3}\frac{1}{8\pi m^2_X}\int\;ds\;\sum_{\substack{{\rm pol}}}|\langle \psi\rho(s)|X\rangle|^2p^*(m^2_{_X},m^2_\psi,s)\frac{1}{\pi}\frac{m_\rho\Gamma_\rho\; \mathcal{B}(\rho\to\pi\pi)}{(s-m_\rho^2)^2+(m_\rho\Gamma_\rho)^2}\frac{m_\rho}{\sqrt{s}}\frac{p^*(s,m^2_{\pi^+},m^2_{\pi^-})}{p^{*}(m_\rho^2,m^2_{\pi^+},m^2_{\pi^-})}.
\ee
Here  $\cal{B}$ denotes a branching fraction, $\Gamma_\rho$ is the width of the $\rho$ resonance, $\langle \psi\rho(s)|X\rangle$ is the transition amplitude of the previous section
\footnote{By $\langle \psi\rho(s)|X\rangle$ we mean $\langle \psi(\epsilon,p)\rho(\eta,q)|X(\lambda,P)\rangle$ with $q^2=s$. $s$ is thus the invariant mass of the $\pi\pi$ pair coming from the $\rho$. 
In what follows we will use the same notation for the transition matrix element to a final state containing an unstable particle.}, and  $p^*$ is the decay momentum in the $X$ rest frame, given by
\be
p^*(x,y,z)=\frac{\sqrt{\lambda(x,y,z)}}{2\sqrt{x}},
\ee
where the K\"{a}ll\'{e}n function is 
\be
\lambda(x,y,z)=x^2+y^2+z^2-2 x y  -2 y z - 2 x z.
\ee

In the calculations we will substitute $m_\rho\Gamma_\rho \to \sqrt{s} \;\Gamma_\rho(s)\to (s/m_\rho) \Gamma_\rho$, the comoving width (see Eq.~(\ref{eq:cw}) in Appendix~\ref{app:rho}).
Similarly for $X\to J/\psi\;\omega\to J/\psi\; \pi^+\pi^-\pi^0$ we get
\begin{eqnarray}
\Gamma(X\to J/\psi\;\pi^+\pi^-\pi^0)&=&\frac{1}{3}\frac{1}{8\pi m^2_X}\int\;ds\;\sum_{\substack{{\rm pol}}}|\langle \psi\;\omega(s)|X\rangle|^2p^*(m^2_X,m^2_\psi,s)\notag\\
&&\times \frac{1}{\pi}\frac{m_\omega\Gamma_\omega\; \mathcal{B}(\omega\to3\pi)}{(s-m_\omega^2)^2+(m_\omega\Gamma_\omega)^2}\frac{\Phi^{(3)}(\sqrt{s},m_{\pi^+},m_{\pi^-},m_{\pi^0})}{\Phi^{(3)}(m_\omega,m_{\pi^+},m_{\pi^-},m_{\pi^0})},
\label{4body}
\end{eqnarray}
where $m_\omega\Gamma_\omega \to (s/m_\omega) \Gamma_\omega$. 
The meaning of $\Phi^{(3)}$ is  explained in Eq.~(\ref{eq:ph3}), Appendix C.

The width of $X\to D^0\bar{D}^{0*}\to D^0\bar{D}^0\pi^0$ can be written as the one for $X\to J/\psi\;\rho\to J/\psi\;\pi^+\pi^-$. 
Using the expressions for the invariant amplitudes in terms of the couplings constants in the preceding section, we
obtain $g_{_{1\psi\rho}} =0.14\pm 0.03$, $g_{_{1\psi\omega}}=0.36\pm 0.01$ and $g_{_{1DD^*}}=(5\pm 1)~{\rm GeV}$.

{\bf The $\bm{J^{PC}=2^{-+}}$ case}. We will use four different couplings to describe the decays 
$X\to J/\psi\;\rho$ and $X\to J/\psi\;\omega$: $g_{_{2\psi\rho}}$, $g_{_{2\psi\rho}}^\prime$ and $g_{_{2\psi\omega}}$, $g_{_{2\psi\omega}}^\prime$.
As for the $J/\psi\;\gamma$ channel, one can assume that the decay proceeds through a hadronic channel: $X$ first decays to $J/\psi\;\rho$ or $J/\psi\;\omega$
and later $\rho$ or $\omega$ convert into a photon, using vector meson dominance
\be
\begin{split}
\langle J/\psi \;\gamma|X\rangle& =\langle \gamma|\omega\rangle\frac{1}{m^2_\omega}\langle J/\psi \;\omega(q^2=0)|X\rangle+\langle \gamma|\rho(q^2=0)\rangle\frac{1}{m^2_\rho} \langle J/\psi \;\rho|X\rangle\\
&=\frac{f_\omega}{m^2_\omega}\langle J/\psi \;\omega(q^2=0)|X\rangle+\frac{f_\rho}{m^2_\rho}\langle J/\psi \;\rho(q^2=0)|X\rangle.
\end{split}
\ee
\noindent We use the decay constants for $\rho$ and $\omega$ derived from the $e^+e^-$ partial decay width of the two mesons: $f_{\rho}=0.121~{\rm GeV}^{2}$ and $f_{\omega}=0.036~{\rm GeV}^{2}$~\cite{Sakurai:1969}.
The matrix element for the decay of $X\to J/\psi\;\gamma$ is thus also written in terms of $g_{_{2\psi\omega}}$, $g_{_{2\psi\omega}}^\prime$ and $g_{_{2\psi\rho}}$, $g_{_{2\psi\rho}}^\prime$. We are left with four couplings to be determined and only three input values for the branching ratios: $\mathcal{B}(X\to\psi\omega)$, $\mathcal{B}(X\to\psi\rho)$ and $\mathcal{B}(X\to\psi\gamma)$.
To perform the fit of the coupling we therefore use the data on the $3\pi$ invariant mass spectrum taken from~\cite{babarbn}. 

\begin{figure}[!h]	
\includegraphics[width=9truecm]{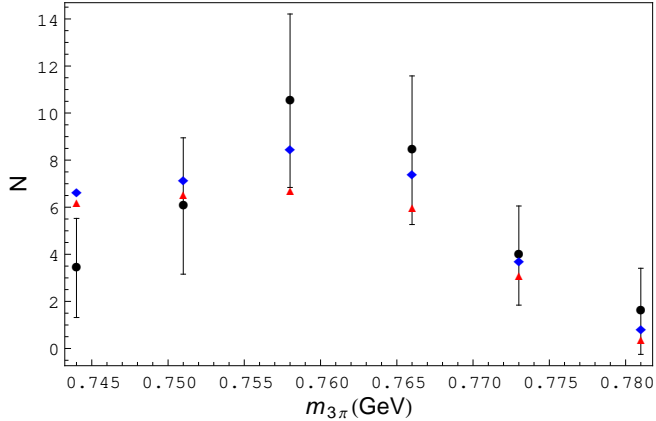}
\caption{Our fit from Eq.~(\ref{prediction}) (Red Triangles) compared with experimental data (Black Disks) 
and the fit in~\cite{babarbn} (Blue Diamonds). $\chi^2/{\rm DOF}=4.03/4$}
\label{fig:fitomegaexp}
\end{figure}

In~\cite{babarbn} $3\pi$ events are selected from a sample of $J/\psi\;\omega$ events with an invariant mass in the interval $3.8625~{\rm GeV}<m_{J/\psi\;\omega}<3.8825~{\rm GeV}$.
To perform the fit we simulate the decay of a $2^{-+}$ particle extracting its squared mass $x_i=m^2_i$ randomly  with a Breit-Wigner distribution centered at $m_{_X}=3.8723~{\rm GeV}$ and with a width $\Gamma_{_X}=0.003~{\rm GeV}$~\cite{SpecNC}.
For each value $x_i$ we require that $x_i>0$ and that $3.8625~{\rm GeV}<\sqrt{x_i}<3.8825~{\rm GeV}$.
Having assigned $m^2_i$ the expected number of $3\pi$ events with a definite invariant mass $m^2_{3\pi}=s$, 
is proportional to the distribution with respect to $s$ of the decay width $\Gamma(X\to J/\psi\;\pi^+\pi^-\pi^0)$
\be
N_i(m^2_{3\pi}=s)\propto \frac{d\Gamma(X\to J/\psi\;\pi^+\pi^-\pi^0)}{ds},
\ee
which can be computed  using Eq.~(\ref{4body}). 
Neglecting the overall numerical normalization we obtain
\be
N_i(m^2_{3\pi}=s)\propto \frac{1}{m^2_i}\sum_{\substack{{\rm pol}}}|\langle J/\psi\;\omega|X(m^2_i)\rangle|^2\frac{1}{(s-m_\omega^2)^2+(\frac{s}{m_\omega}\Gamma_\omega)^2}p^*(m^2_i,m^2_\psi,s) \Phi^{(3)}(\sqrt{s},m_\pi^+,m_\pi^-,m_\pi^0)
\ee
\noindent  if $m_i>m_{\psi}+\sqrt{s}$. 
Thus the total number of events at fixed $s$ is
\be
\label{prediction}
N(m^2_{3\pi}=s)=\sum_{i}N_{i}(m^2_{3\pi}=s)\;\theta(m_i-m_\psi-\sqrt{s}).
\ee
In Fig.~\ref{fig:fitomegaexp} we show the agreement obtained with data ($\chi^2/{\rm DOF}=4.03/4$) and  we compare it with the experimental fit obtained using a Blatt-Weisskopf factor to account for the $\ell=1$ decay, as was done in~\cite{babarbn}.

To compute the normalization factor we exploit the partial
decay width of $\Gamma(X\to\psi\omega)=\mathcal{B}(X\to\psi\omega)\Gamma_X$ as written in Eq.~(\ref{4body}).
We obtain $g_{_{2\psi\omega}}=(1.58\pm0.16)~{\rm GeV}^{-1}$ and $g_{_{2\psi\omega}}^\prime=(-0.74\pm0.34)~{\rm GeV}^{-1}$. 
Using the known experimental data on $\mathcal{B}(X\to\psi\rho)$ and $\mathcal{B}(X\to\psi\gamma)$ we obtain two possible solutions for $g_{_{2\psi\rho}}$ and $g^\prime_{_{2\psi\rho}}$.
Since the $J/\psi\rho\to X_2\to D^0\bar{D}^{0*}$ cross-section turns out to be roughly the same using the two sets of couplings, we choose to use one of them, namely $g_{_{2\psi\rho}}=(-0.29\pm0.08)~{\rm GeV}^{-1}$, $g_{_{2\psi\rho}}^\prime=(0.28\pm0.09)~{\rm GeV}^{-1}$.

If one fits the same data set assuming $J^{PC}=1^{++}$ a $\chi^2/{\rm DOF}\sim 9/4$ is obtained,
which means that the probability of the $1^{++}$ hypothesis is smaller by a factor of $6$ than the $2^{-+}$ one.
For the decay $X\to D^0\bar{D}^{0*}$ we use the same method to extract the coupling and we obtain $g_{_{2DD^*}}=267\pm50$.

The results are summarized in Table~{\ref{tab:couplings}}.
As a consequence of the fact that $\mathcal{B}(X\to D^0\bar{D}^{0*})>\mathcal{B}(X\to \psi\omega)$ and $\mathcal{B}(X\to D^0\bar{D}^{0*})>\mathcal{B}(X\to \psi\rho)$
we find that the adimensional coupling $g_{_{2DD^*}}$ is much larger than $g_{_{1\psi\omega}}$ and $g_{_{1\psi\rho}}$.
On the other hand all the dimensional couplings turn out to be of the same order of magnitude of the mass scales involved.

\begin{table}[ht!]
\begin{tabular}{||c|c|l|l||} \hline
\backslashbox{{\rm coupling}}{$J^{PC}$} & $1^{++}$ & \multicolumn{2}{|c||}{$2^{-+}$} \\ \hline
$g_{_{(J)DD*}}$ & $(5\pm 1)~{\rm GeV}$ & \multicolumn{2}{|c||}{$267\pm50$} \\ \hline
$g_{_{(J)\omega\psi}}$ & $0.36\pm 0.01$  & $1.58\pm0.16~{\rm GeV}^{-1}$ & $-0.74\pm0.34~{\rm GeV}^{-1}$\\ \hline
$g_{_{(J)\rho\psi}}$ & $0.14\pm 0.03$ & $(-0.29\pm0.08)~{\rm GeV}^{-1}$ & $(0.28\pm0.09)~{\rm GeV}^{-1}$\\\hline
\end{tabular}
\caption{Fitted values for the effective couplings of $X(3872)$ to $D\bar{D}^{0*}$, $J/\psi\;\omega$ and $J/\psi\;\rho$ for the two $J^{PC}$ assignments.}
\label{tab:couplings}
\end{table}

\section{ An application to the $J/\psi$ Suppression in Hot Hadronic Matter}
Recently the PHENIX collaboration published new data on the $J/\psi$ suppression in heavy ion collisions observed at RHIC~\cite{Adare:2006ns},
which have been discussed for example in~\cite{Capella:2007jv}. 
These data, together with the upcoming ones from the LHC - ALICE collaboration, have encouraged us to 
consider, as a possible application of the determination of the $X_{1,2}$ coupling strengths, the   
study of the contribution 
of the $X(3872)$ to the $J/\psi$ suppression by a hot hadron gas and to revise some previous results on this topic.

A decrease of the $J/\psi$ production in heavy ion collisions is one of the 
first quark gluon plasma discovery signals suggested in the literature~\cite{Matsui:1986dk}.
However processes like 
\be
\label{standard}
J/\psi\;(\pi,\eta,\rho,\omega, \phi,K^{(*)}, ...)\to D^{(*)}\bar{D}^{(*)},
\ee 
\noindent which are at work in a hypothetical hadron gas formed in place or after the deconfined phase of quarks and gluons, may also 
provide a source of attenuation of $J/\psi$ - an antagonist signal to the standard one of quark gluon plasma suppression.
These contributions might also take place at  a different stage of the hadronization process -- once the plasma has converted into hadrons under the hypothesis that the hadron gas is itself in thermal equilibrium. 
Such situations have been extensively studied in the past. Here we  take into account also the in-medium effects on the open 
charm mesons discussed in~\cite{Faessler:2002qb,Fuchs:2004fh,Tolos:2007qr,Tolos:2007vh,Gottfried:1992bz,Hayashigaki:2000es,Kumar:2009xc,Kumar:2010gb,Kumar:2010nz} and update analyses such as those in~\cite{Maiani:2004py,Maiani:2004qj,Wong:1999zb,Wong:2001td,Armesto:1997sa,Armesto:1998rc,Capella:2005cn} and lattice studies like~\cite{Yokokawa:2006td}. 
Similar studies can be found in~\cite{Sibirtsev:1999jr} or in~\cite{Burau:2000pn,Blaschke:2002ww}, where a critical temperature (Mott transition) is introduced to reproduce the dip observed in $J/\psi$ suppression in correspondence of a particular centrality; see also \cite{Wong:2001uu,Xu:2002zv,Morita:2010pd}.

In contrast to previous works, we consider here a resonant channel mediated by $X_{1,2}$ which turns out to be relevant if  in-medium effects on open charm mesons are considered.
Due to the narrowness of the $X(3872)$, one would expect the contribution of the {\it s}~-~channel processes
\be
\label{xx}
J/\psi\;(\rho,\omega)\to X(3872)\to D^{0}\bar{D}^{0*}
\ee
\noindent to be negligible. 
Nevertheless they can be enhanced because the properties of open charm mesons change when propagating inside a hadron medium. 
In particular their masses are expected to decrease, lowering the $D^0\bar{D}^{0*}$ threshold.
Also the non-resonant modes are affected by in-medium $D$ meson properties. Hence we re-analyze some results previously obtained~\cite{Maiani:2004py,Maiani:2004qj,Yokokawa:2006td,Wong:1999zb,Wong:2001td}, 
in particular those in~\cite{Maiani:2004qj}.

We will obtain an estimate for the cross-section for the process of Eq.~(\ref{xx}), using the couplings derived in the first part of this paper.
In the following we will briefly review how the properties of open charm mesons are expected to be modified inside the hadron medium.
Using the results found in the literature we will quantify the effect of this dissociation process
and update the estimates on the non resonant channels of Eq.~(\ref{standard}) with respect to 
those given in~\cite{Maiani:2004qj}.
We will compare the predictions obtained  with the experimental data in the last subsection of the paper. 

\subsection{Cross-sections}
\label{subsec:xsect}

\begin{figure}[!h]
\begin{center}
\begin{minipage}[t]{7truecm} % A minipage that covers half the page
\centering
\includegraphics[width=7.5truecm]{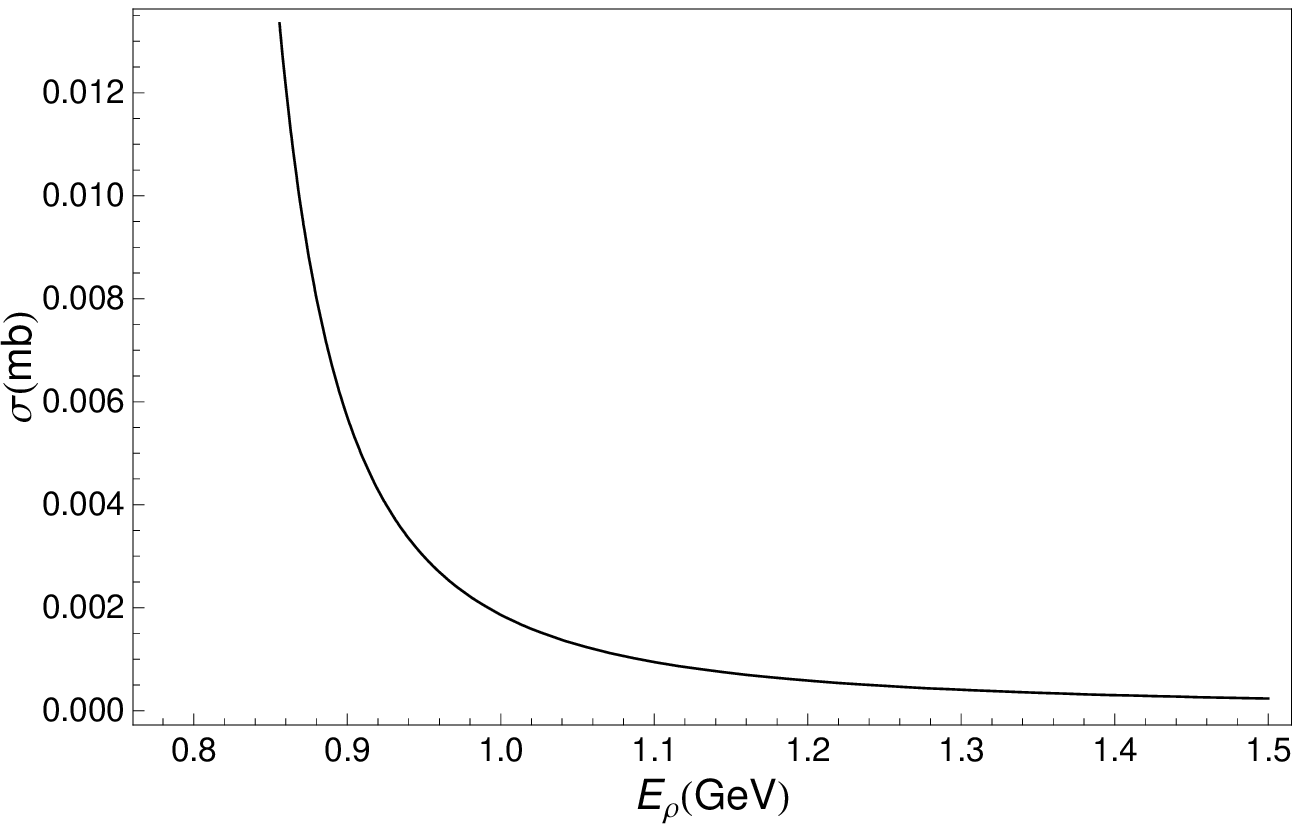}
\caption{Dissociation cross-section of $J/\psi$ into open charm mesons mediated by $X_1$ as a function of the energy of the $\rho$ in the rest frame of the $J/\psi$ ($g_{_{1\psi\rho}}=0.14$, $g_{_{1DD^*}}=5~{\rm GeV}$). The low energy `exothermic' peak is present.}
\label{fig:1pprho}
\end{minipage}
\hspace{1truecm} % To get a little bit of space between the figures
\begin{minipage}[t]{7truecm}
\centering
\includegraphics[width=7.5truecm]{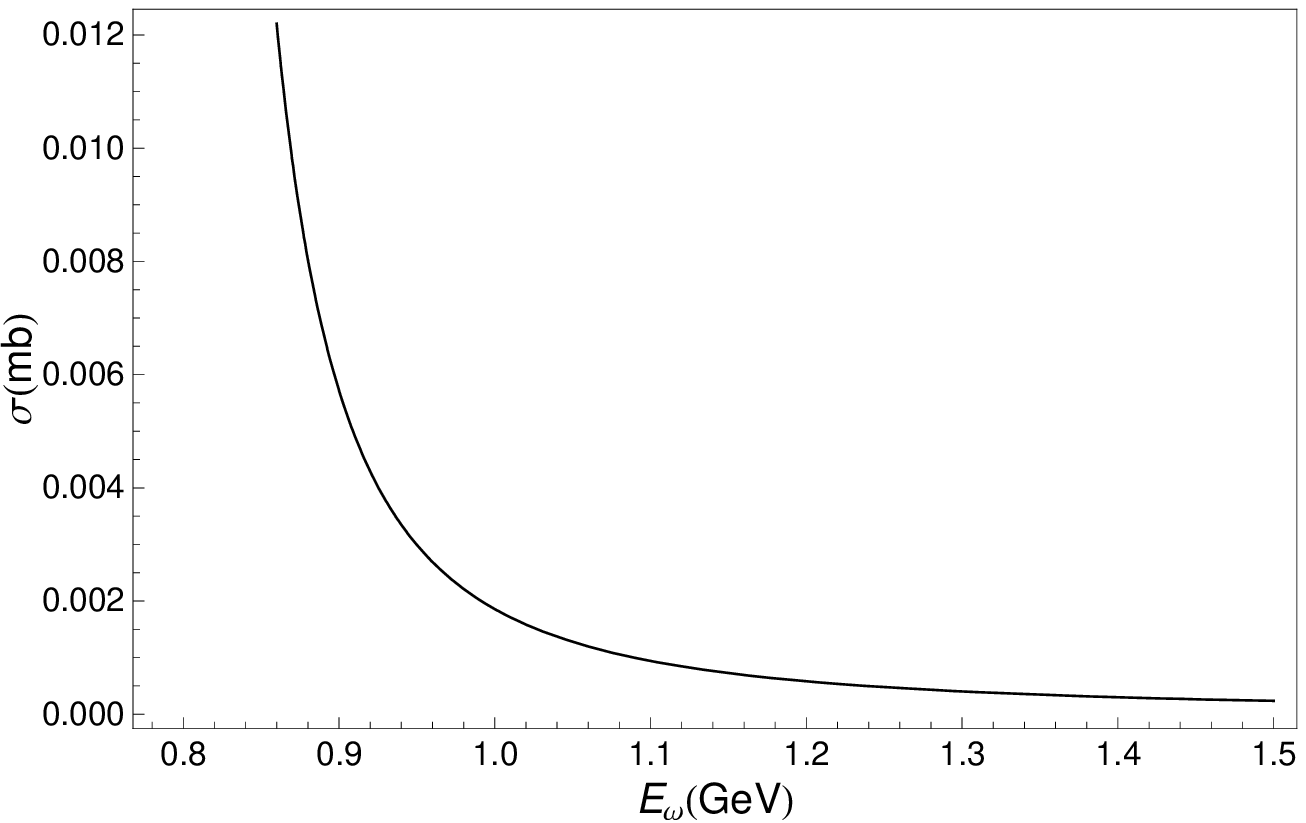}
\caption{Dissociation cross-section of $J/\psi$ into open charm mesons mediated by $X_1$ as a function of the energy of the $\omega$ in the rest frame of the $J/\psi$ ($g_{_{1\psi\omega}}=0.36$, $g_{_{1DD^*}}=5~{\rm GeV}$). The low energy `exothermic' peak is present.}
\label{fig:1ppomega}
\end{minipage}
\end{center}
\end{figure}

\begin{figure}[!h]
\begin{center}
\begin{minipage}[t]{7truecm} % A minipage that covers half the page
\centering
\includegraphics[width=7.5truecm]{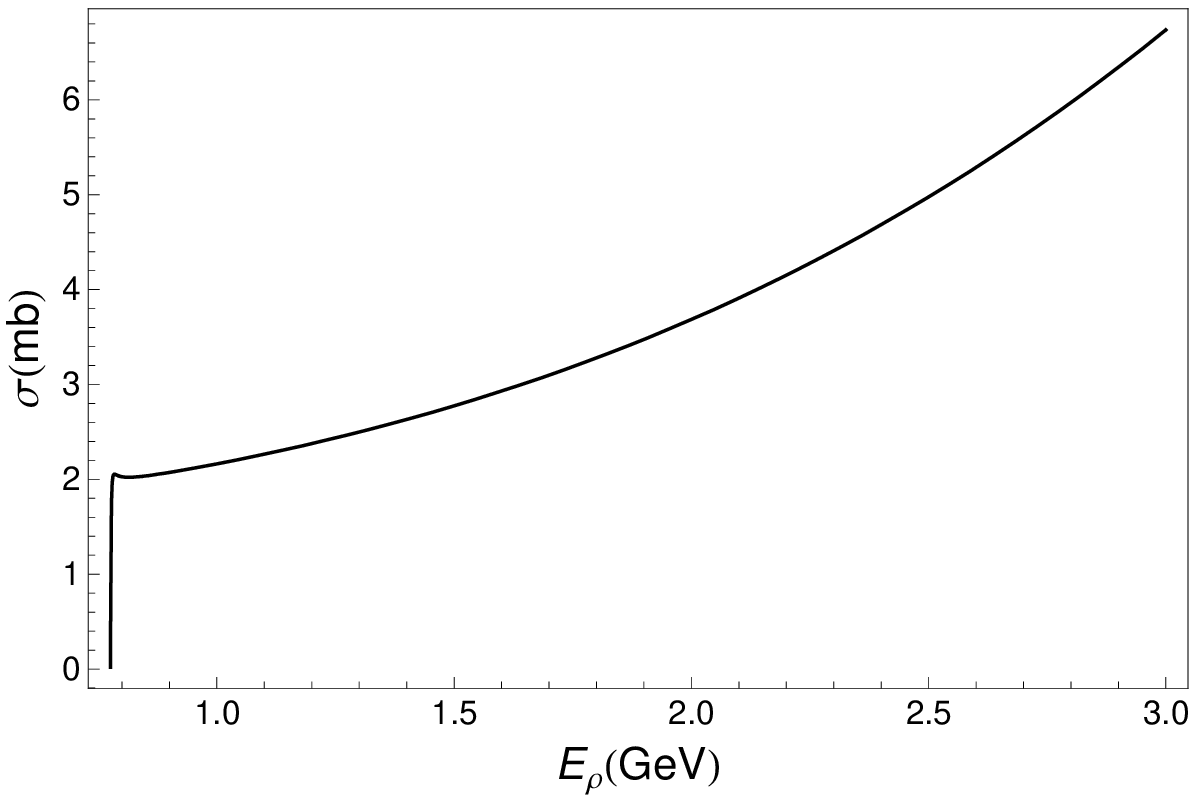}
\caption{Dissociation cross-section of $J/\psi$ into open charm mesons mediated by $X_2$ as a function of the energy of the $\rho$ in the rest frame of the $J/\psi$ ($g_{_{2\psi\rho}}=-0.29~{\rm GeV}^{-1}$, $g^\prime_{_{2\psi\rho}}=0.28~{\rm GeV}^{-1}$ and $g_{_{2DD^*}}=267\pm50$). If one uses the other set of couplings for $\rho$ the cross-section is roughly the same.}
\label{fig:2mprho}
\end{minipage}
\hspace{1truecm} % To get a little bit of space between the figures
\begin{minipage}[t]{7truecm}
\centering
\includegraphics[width=7.5truecm]{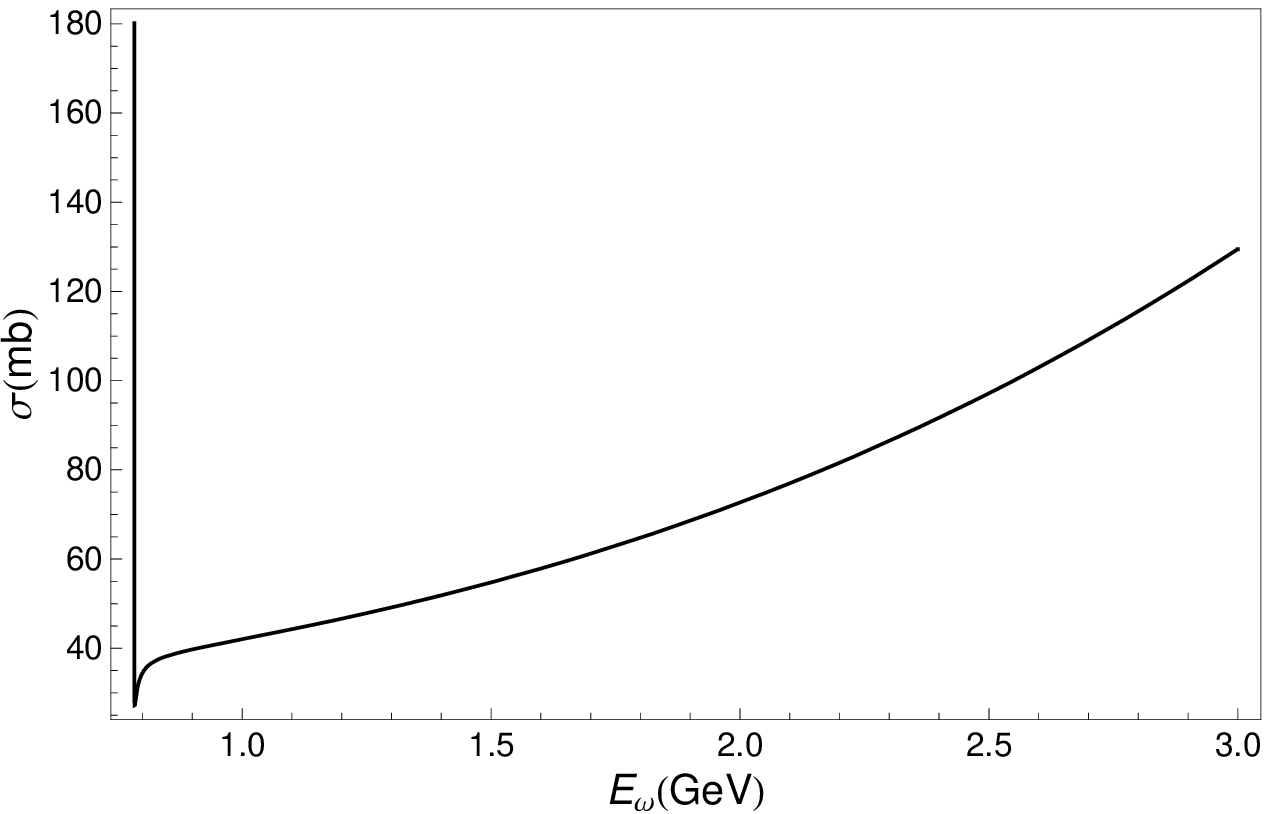}
\caption{Dissociation cross-section of $J/\psi$ into open charm mesons mediated by $X_2$ as a function of the energy of the  $\omega$ in the rest frame of the $J/\psi$ ($g_{_{2\psi\omega}}=1.58~{\rm GeV}^{-1}$, $g_{_{2\psi\omega}}^\prime=-0.74~{\rm GeV}^{-1}$  and $g_{_{2DD^*}}=267$). Consider that we are actually concerned only with relatively low energy $\rho$ and $\omega$ mesons in a Hagedorn gas.}
\label{fig:2mpomega}
\end{minipage}
\end{center}
\end{figure}

The cross-section for the process of Eq.~(\ref{xx}), depicted in Fig.~\ref{fig:graph}, reads as follows~(see~Appendix \ref{app:xsect})
\be
\label{xsect}
\sigma(J/\psi\;\rho\to D\bar{D}^{0*})=\frac{1}{9}\frac{1}{2\;\lambda^{1/2}(s,m^2_\psi,m^2_\rho)}	\;f(s,m^2_\psi,m^2_\rho,m^2_{{X_{1,2}}},m^2_{_D},m^2_{_{D^*}})\;\frac{1}{16\pi}\frac{\lambda^{1/2}(s,m^2_{_D},m^2_{_{D^*}})}{s},
\ee
\noindent where $f(s)$ is the integral over the scattering angle of the sum over polarizations of the squared matrix element 
\be
f(s,m^2_\psi,m^2_\rho,m^2_{{X_{1,2}}},m^2_{_D},m^2_{_{D^*}})=\int d\cos\theta\sum_{{\rm pol}}|\M_{{\rm via}\;X_{1,2}}|^2(s,m^2_\psi,m^2_\rho,m^2_{{X_{1,2}}},m^2_{_D},m^2_{_{D^*}},\theta).
\ee
For the matrix elements we use the couplings reported in Table~\ref{tab:couplings}.
The resulting cross-sections are shown in Fig.~\ref{fig:1pprho}-\ref{fig:2mpomega} as  functions of $E_\rho$ or $E_\omega$, the energies of the $\rho$ and the $\omega$ in the rest frame of the $J/\psi$:
$s = m^2_{\rho,\omega}+m^2_\psi+2m_\psi E_{\rho,\omega}$.

The functional behavior of the cross-sections shown can be explained as follows. 
At small values of the energy of the incoming $\rho,\omega$ the ``exothermic" peak appears\footnote{In Fig.~\ref{fig:2mprho} the peak is not resolved because of the $x$-scale chosen.}:
the threshold energy of the reaction $m_{_D}+m_{_{D^*}}$ is indeed smaller than the minimum value of $\sqrt{s}$, namely $m_\rho+m_\psi$, 
so that the divergence in the flux factor is located at a larger value than the threshold one.
At higher energies, $s>>m^2_{X}$, the flux factor  behaves as $1/s$ , whereas the phase space is approximately constant ($\lambda^{1/2}(s,0,0)/s\simeq 1$) so that
\be
\sigma(s)\sim\frac{1}{s}\times f(s)\;\;\;{\rm as} \;\; s>>m^2_X.
\ee
Here comes the difference between the $1^{++}$ and the $2^{-+}$ assignments.
In the $X_1$ case at high energies $f(s)\sim s^0$ giving $\sigma(s)\sim 1/s$.
If instead $X=X_2$, $f(s)\sim s^7$ giving instead $\sigma(s)\sim s^6$.

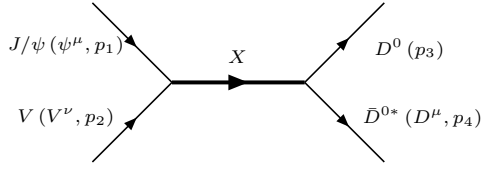
\begin{figure}[!h]
\begin{center}
\begin{picture}(150,100) (0,0)
\SetWidth{0.6}
\ArrowLine(20,20)(50,50)
\ArrowLine(20,80)(50,50)
\SetWidth{1.5}
\ArrowLine(50,50)(100,50)    
\SetWidth{0.6}
\ArrowLine(100,50)(130,80)
\ArrowLine(100,50)(130,20)
\Text(10,64)[]{\scriptsize{$J/\psi\left(\psi^\mu,p_1\right)$}}
\Text(10,37)[]{\scriptsize{$V \left(V^\nu,p_2\right)$}}
\Text(140,63)[]{\scriptsize{$D^0\left(p_3\right)$}}
\Text(145,37)[]{\scriptsize{$\bar{D}^{0*}\left(D^\mu,p_4\right)$}}
\Text(75,60)[]{\scriptsize{$X$}}
\end{picture}
\end{center}
\caption{Feynman graph for $J/\psi\;(\rho,\omega)\to  X(3872)\to D^0\bar{D}^{0*}$.}
\label{fig:graph}
\end{figure}

\subsection{In-medium properties of open charm mesons}
\label{subsec:medium}
The modifications of the masses and decay widths of open charm mesons $D^0$ and $D^{0*}$ inside
a hot pion gas have been computed for example in~\cite{Fuchs:2004fh}, following the approach discussed in~\cite{kapusta}. 
Indeed the presence of a gas of light hadrons, such as $\pi$'s, 
can sustain scattering processes which involve $D$-mesons leading to a modification of their masses and widths.
These two quantities are both related to the self-energy diagrams, which can be written at finite temperature as the thermal averages of the 
resonant part of the $D\pi^{\pm,0}$ forward scattering amplitude.

\begin{figure}[!h]
\begin{center}
\begin{minipage}[t]{7truecm} % A minipage that covers half the page
\centering
\includegraphics[width=7.5truecm]{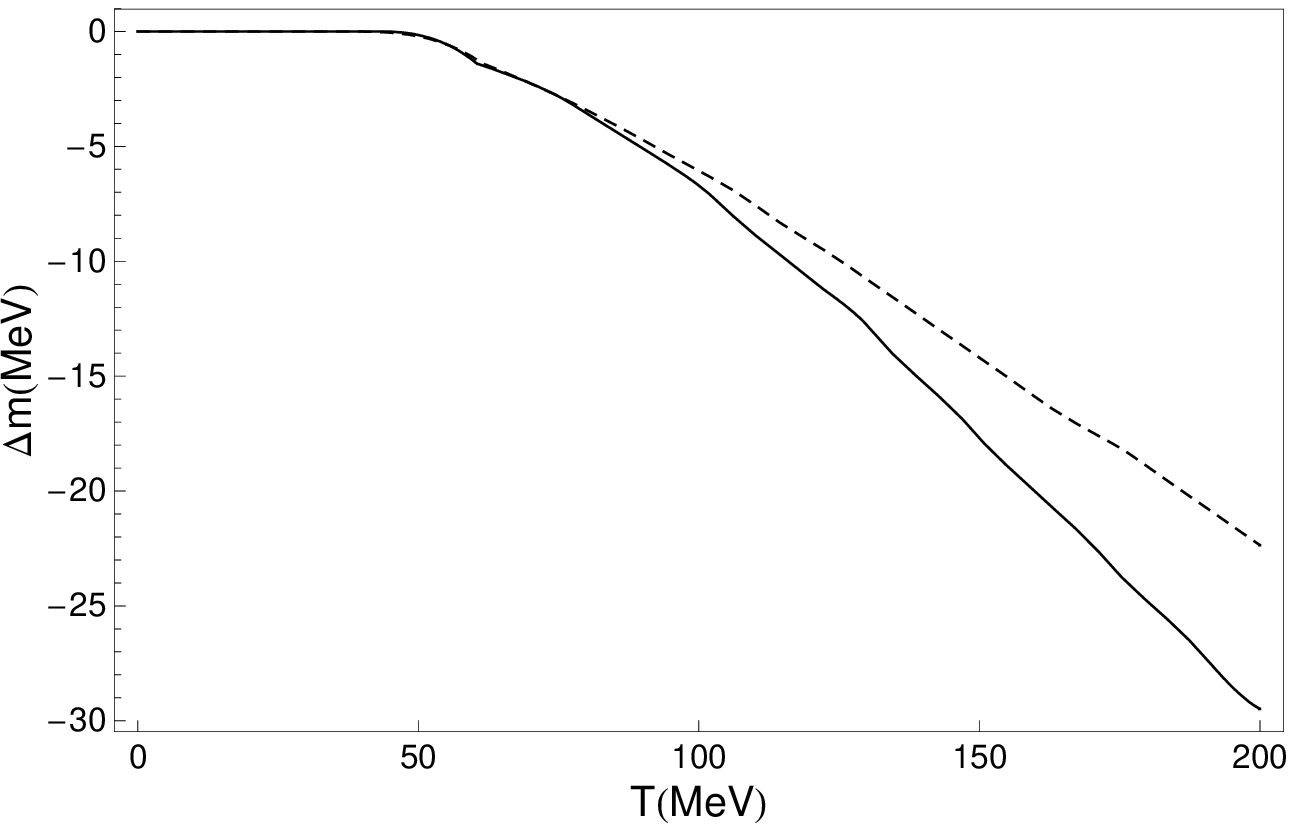}
\caption{In medium mass modification computed in~\cite{Fuchs:2004fh} for $D^0$ (solid line) and $D^{0*}$ (dashed line).}
\label{fig:deltam}
\end{minipage}
\hspace{1truecm} % To get a little bit of space between the figures
\begin{minipage}[t]{7truecm}
\centering
\includegraphics[width=7.5truecm]{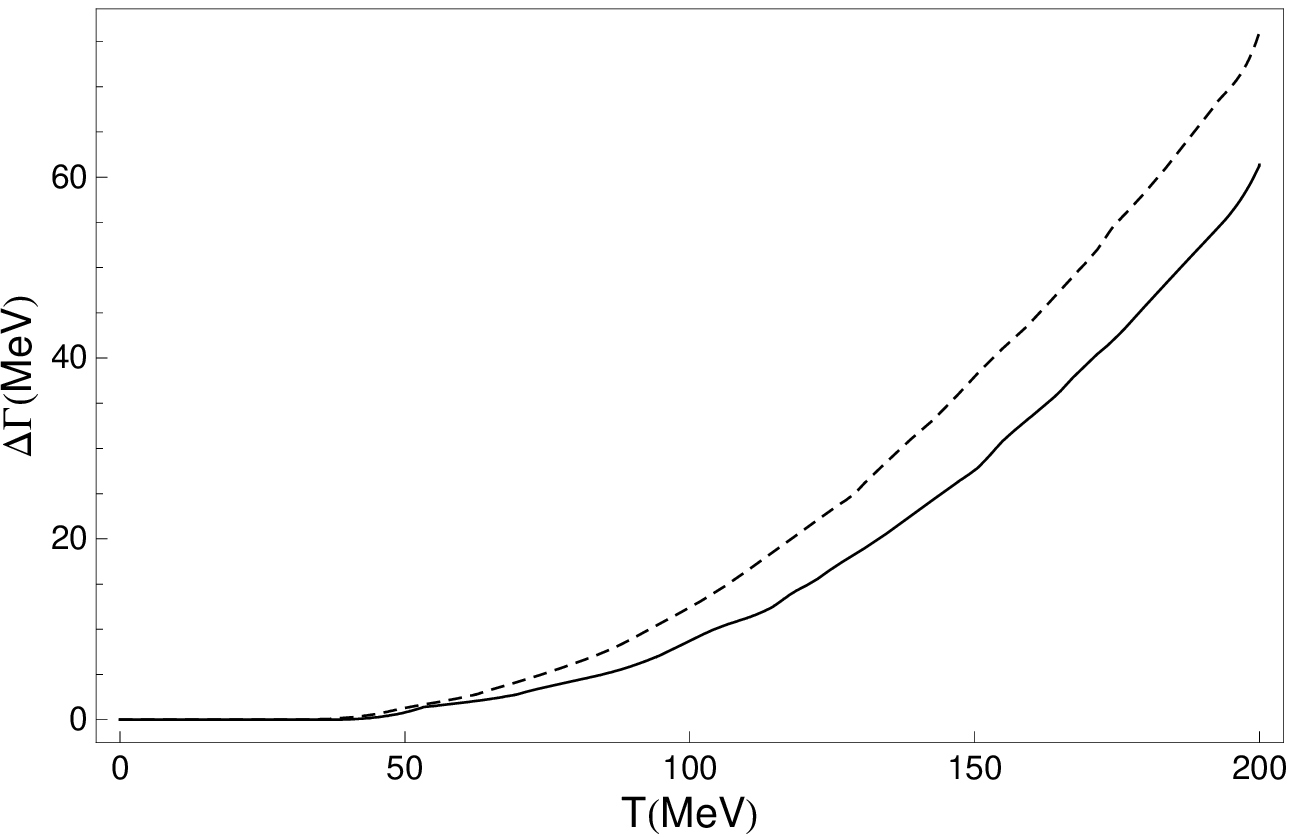}
\caption{In medium total decay width computed in~\cite{Fuchs:2004fh} for $D^0$ (solid line) and $D^{0*}$ (dashed line).}
\label{fig:deltagamma}
\end{minipage}
\end{center}
\end{figure}

The decrease of the mass and the increasing decay width for both the $D$-mesons found in~\cite{Fuchs:2004fh} are shown in Fig.~\ref{fig:deltam} and Fig.~\ref{fig:deltagamma}.

The authors of~\cite{Burau:2000pn,Blaschke:2002ww}, obtained similar results but with a different approach.
They assume that the shape of $q'\bar{q}$ interaction potentials, responsible for the binding of mesons,
is sensitive to the temperature. Thus it can happen that some discrete levels, corresponding to different $c\bar{q}$
excitations, are shifted into the continuous part of the spectrum becoming metastable states with different masses and non-vanishing widths.
Each $D$-meson excitation undergoes this transition at a different critical temperature: $\Delta M\propto -(T-T_{_C})\theta(T-T_{_C})$.

Since we do not find any relevant differences on the $J/\psi$ dissociation by using the two approaches, we will consider only the one in~\cite{Fuchs:2004fh}.

\begin{figure}[!h]	
\includegraphics[width=9truecm]{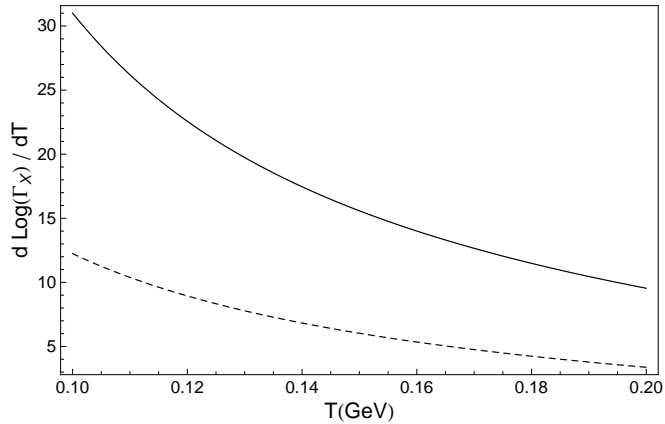}
\caption{Logarithmic derivative of the total decay width of $X(3872)$ as a function of the temperature in the case $X=X_1$ and $m_{_X}=m_{_D}(T)+m_{_{D^*}}(T)$ (dashed line) and $X=X_2$ and $m_{_X}=3872~{\rm MeV}$ (solid line).}
\label{fig:gx}
\end{figure}

The broadening and shifting of the masses of the two open charm mesons lead to a modification of the decay width and mass of the $X(3872)$. 
Since we do not have clues on how in-medium effects would modify the mass of a $X_{1,2}$ tetraquark, we simply assume that 
if $X(3872)=X_1$ it is a $D^{0}\bar{D}^{0*}$ molecule; if $X(3872)=X_2$ it is a charmonium state.
We remind here that the $1^{++}$ assignment is severely at odds with a $2\;^3P_1$ standard charmonium interpretation essentially because 
of the small radiative transition rate $X\to J/\psi\;\gamma$ with respect to what expected. 
In the molecular interpretation the mass of the $X_1$
is directly related to the sum of the masses of the $D^0$ and $D^{0*}$ and thus it will decrease with the temperature.
In the charmonium assignment (and likely also for tetraquarks) one might expect the mass of the $X_2$ to be almost stable with temperature.
This is because $X_2$ would be the $1^1D_2$ charmonium radial ground state and Debye screening is not expected to alter the lowest lying levels~\cite{Matsui:1986dk}.
The $D^0\bar{D}^{0*}$ width  can be computed as

\be
\begin{split}
\Gamma(X\to D^0\bar{D}^{0*})_{_T}&=\frac{1}{2s_{_X} +1}\frac{1}{8\pi m^2_{_X}}\;\int_{s^{{\rm min}}_2}^{s^{{\rm max}}_2}\;ds_2\int_{s^{{\rm min}}_1}^{s^{{\rm max}}_1}\;ds_1\sum_{{\rm pol}}|\M_{X\to D D*}(s_1,s_2)|^2\;\frac{\sqrt{\lambda(m^2_X,s_1,s_2)}}{2m_{_X}}\\
&\times BW(s_1,m_{_D}(T),\Gamma_{_D}(T))\;\;BW(s_2,m_{_{D^*}}(T),\Gamma_{_{D^*}}(T)),
\end{split}
\ee
where by $BW$ we mean the standard normalized Breit-Wigner function
\be
 BW(s,m,\Gamma)=\frac{1}{\pi}\frac{m\Gamma}{(s-m^2)^2+(m\Gamma)^2}
\ee
and $s^{{\rm min}}$ and $s^{{\rm max}}$ are fixed by the kinematics.
We show the results in terms of the logarithmic derivative of the total width of $X_{1,2}$ with respect to the temperature, see Fig.~\ref{fig:gx}: 
in-medium effects make the $X_2$ become much broader than $X_1$.
This fact can be understood by taking into account the dependence of the decay width on the masses of the particles in the final state.
The phase space volume is enlarged proportionally to the decay momentum $p^*$. As for the matrix element,
if $J^{PC}=1^{++}$ the $X\to D^0\bar{D}^{0*}$ decay has $\ell=0$ and thus $|\mathcal{M}|^2\sim{\rm constant}$, 
while if $X$ is a $2^{-+}$ state it has $\ell=1$ so that $|\mathcal{M}|^2\propto p^{*2}$. 
Thus if $J^{PC}=1^{++}$ then $\Gamma_X\propto p^*$, instead if $J^{PC}=2^{-+}$ then $\Gamma_X\propto p^{*3}$.

To summarize, the fact that the charmonium $X_2$ mass is not affected by the medium makes the $X_2\to D^0\bar{D}^{0*}$ {\it P}~-~wave decay
much larger because the $D^0$ and $D^{0*}$ masses are instead sensitively decreased in the finite temperature medium.

\subsection{Comparison to data on $J/\psi$ suppression at RHIC}
\label{subsec:data}

The average absorption length (mean free path) of the $J/\psi$ due to the presence of a $\rho$ meson gas at temperature $T$ is the inverse of the thermal average of the product of  the density number $\rho$ of $\rho$ mesons and the cross-section $\sigma$ (given  in Eq.~(\ref{xx})):
\be
\label{average}
\begin{split}
\langle \rho\sigma_{J/\psi\;\rho\to D^0\bar{D}^{0*}}\rangle_{_T}&=(2s_\rho+1)\int \frac{d^3 p_\rho}{(2\pi)^3}\frac{\sigma(E_\rho)}{e^{E_\rho/\kappa_B T}-1}=\frac{2s_\rho+1}{2\pi^2}\int_{E^{{\rm min}}_\rho}^{E^{{\rm max}}_\rho}\,dE_\rho\frac{p_\rho\,E_\rho\,\sigma(E_\rho)}{e^{E_\rho/\kappa_B T}-1}.
\end{split}
\ee
\noindent Here the kinematics imposes that
\be
E^{{\rm min}}_\rho={\rm max}\left[m_\rho,\sqrt{\frac{(m_{_D}+m_{_{D^*}})^2-m^2_\rho-m^2_\psi}{2m^2_\psi}}\right].
\ee
By numerical inspection we have found that it is safe to cut-off the integrals at
$E^{{\rm max}}_{\rho,\omega}=1.5~{\rm GeV}$ and $E^{{\rm max}}_{\rho,\omega}=3.5~{\rm GeV}$ for $J^{PC}=1^{++}$ and $J^{PC}=2^{-+}$ respectively.
The difference between the two values for $E^{{\rm max}}$ can be understood noting that the cross-section diminishes as the energy grows if $J^{PC} = 1^{++}$, while it grows with energy for $J^{PC} = 2^{-+}$.

Given that the masses of the $D^0$ and $D^{0*}$ mesons are supposed to change with the temperature we need to take into account this effect in the calculation of the thermal averages. 
We average the absorption length over the Breit-Wigner distributions of the $D$ and of the $D^*$: the formula for $\langle \rho\sigma\rangle_T$ is therefore

\be
\label{length}
\begin{split}
\langle \rho\sigma_{J/\psi\;\rho\to X_{1,2}\to DD^*}\rangle_{_T} &=\frac{2s_\rho+1}{2\pi^2}\;\int_{s^{{\rm min}}_2}^{s^{{\rm max}}_2}\;ds_2\int_{s^{{\rm min}}_1}^{s^{{\rm max}}_1}\;ds_1\int_{E^{{\rm min}}_{\rho}}^{E^{{\rm max}}_{\rho}}\,dE_\rho\frac{p_\rho\,E_\rho\,\sigma(E_{\rho},s_1,s_2,m_{_X}(T),\Gamma_{_X}(T))}{e^{E_{\rho}/\kappa_B T}-1}\\
&\times BW(s_1,m_{_D}(T),\Gamma_{_D}(T))\;\;BW(s_2,m_{_{D^*}}(T),\Gamma_{_{D^*}}(T)).
\end{split}
\ee
As already mentioned in the previous subsection, we report only the results obtained using the masses and widths of the $D$-mesons computed in~\cite{Fuchs:2004fh}. 
If one uses the discontinuous functions for $m_{_D}(T)$ and $\Gamma_{_D}(T)$ proposed
in~\cite{Burau:2000pn,Blaschke:2002ww}, 
the values obtained for $\langle \rho\sigma\rangle_{_T}$ are of the same magnitude.
Moreover, regardless of whether the non-resonant channel is included, $\langle \rho\sigma\rangle_{_T}$ does {\it not} show any discontinuity that can help in fitting the observed dip in the experimental data, contrarily to what shown in~\cite{Burau:2000pn,Blaschke:2002ww}. 
The same holds for the $\omega$.

\begin{figure}[!h]
\begin{center}
\begin{minipage}[t]{7truecm} % A minipage that covers half the page
\centering
\includegraphics[width=8truecm]{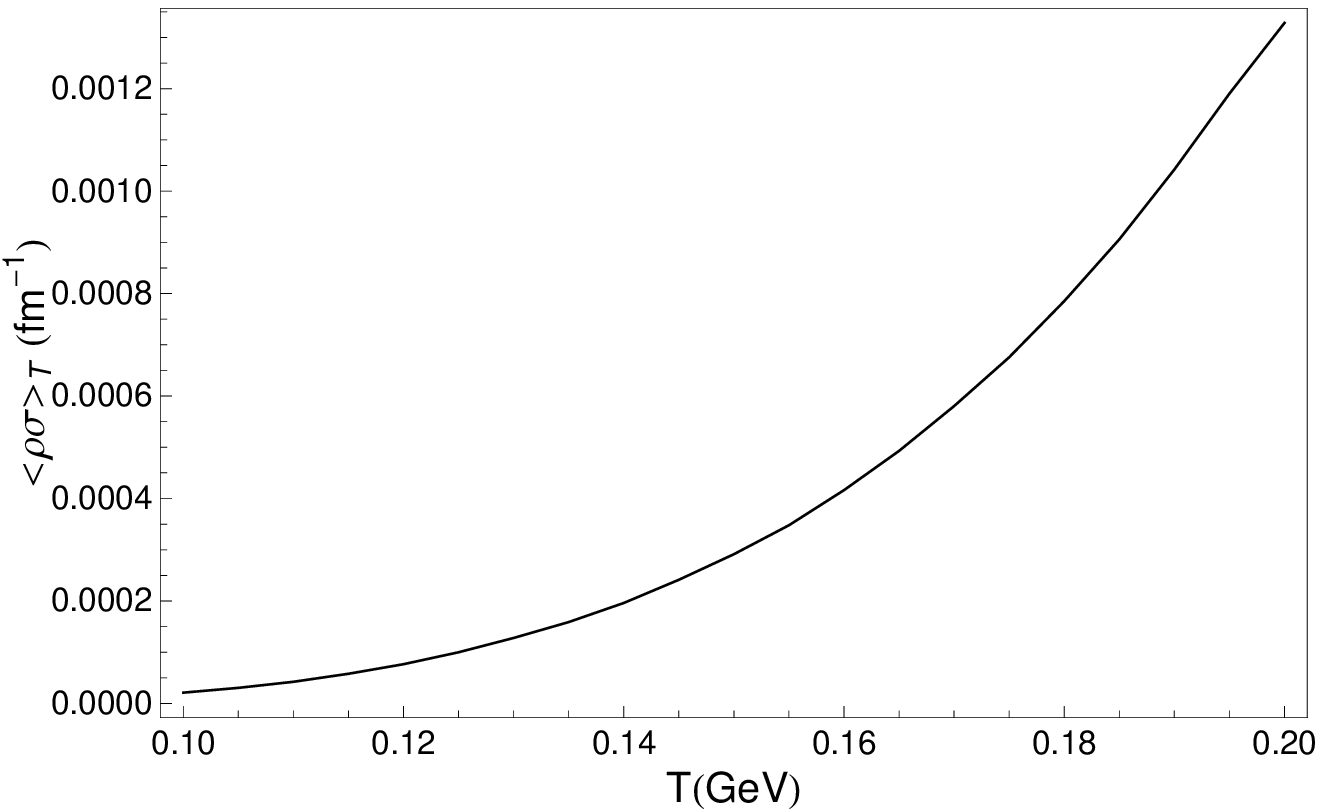}
\caption{$J/\psi\;\rho\to D^0\bar{D}^{0*}$: Inverse average absorption length for $X_2$ hypothesis, using $m_{_X}=3.8723~{\rm GeV}$.}
\label{fig:rsigma2mprho}
\end{minipage}
\hspace{1truecm} % To get a little bit of space between the figures
\begin{minipage}[t]{8truecm}
\centering
\includegraphics[width=8truecm]{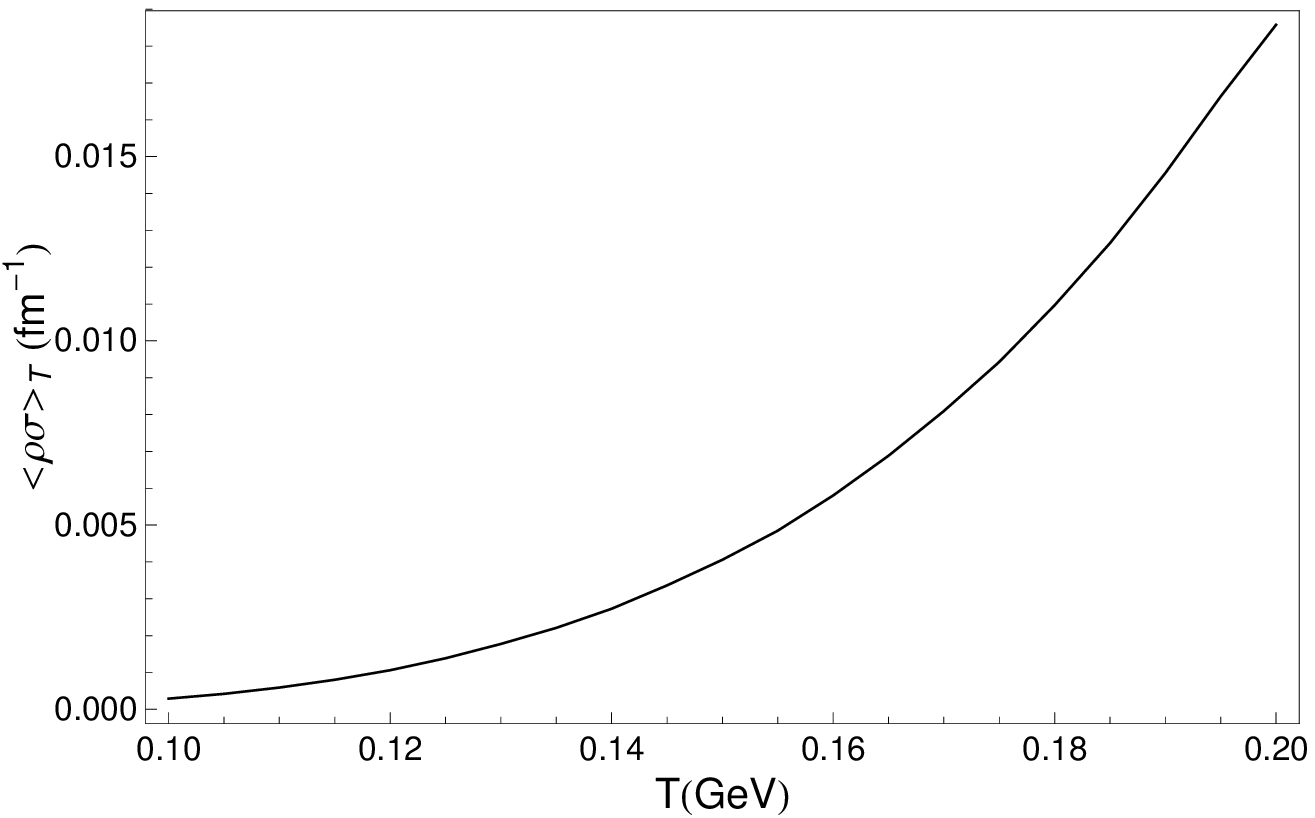}
\caption{$J/\psi\;\omega\to D^0\bar{D}^{0*}$: Inverse average absorption length for $X_2$ hypothesis, using $m_{_X}=3.8723~{\rm GeV}$.}
\label{fig:rsigma2mpomega}
\end{minipage}
\end{center}
\end{figure}

In Fig.~\ref{fig:rsigma2mprho} and \ref{fig:rsigma2mpomega} we show the results for the inverse average absorption length for the resonant $J/\psi$ suppression mediated by $X_2$ and initiated by $\rho$ and $\omega$ respectively. 
For the $X_1$ case we find the effect is negligible, since the in-medium $X_1$ is still too narrow for $\rho$ and $\omega$ to effectively dissociate the $J/\psi$ into open charm mesons.

We also update the estimates for the non-resonant channels enumerated in Eq.~(\ref{standard}) as discussed in~\cite{Maiani:2004qj}.
In Sec.~\ref{app:multiplicitynonres} of the Appendix, we give some details on the counting rules for all the $J/\psi$ absorption processes we consider in the hadron gas.

In Table~\ref{tab:absorption} we give a summary of the results for the inverse mean free paths. 
The contribution of the $X_1$ is negligible whereas the contribution from the $X_2$ resonant channel is of the same size of the sum of the non-resonant channels up to about $T=120$~MeV. With the growing temperature the resonant contribution is found to weight less than the non-resonant ones, reducing to a $15\%$ of the non-resonant total at about  the Hagedorn temperature  $T\sim 170$~MeV.
We remind the reader that we have neglected possible interference between resonant and non resonant channels.

\renewcommand{\arraystretch}{1.0}
\begin{center}
\begin{table}[ht!]
\begin{tabular}{||c|c|c|c|c|c|c|c|c|c||}
\hline
$T$~(GeV) & $\langle \rho\sigma\rangle_T^{(\rho+\omega)_{_{X_2}}}$~(fm$^{-1}$) &$\langle \rho\sigma\rangle_T^{\rho+\omega}$~(fm$^{-1}$) & $\langle \rho\sigma\rangle_T^{K}$~(fm$^{-1}$) & $\langle \rho\sigma\rangle_T^{K^*}$~(fm$^{-1}$) & $\langle \rho\sigma\rangle_T^{\phi}$~(fm$^{-1}$) &$\langle \rho\sigma\rangle_T^{\pi+\eta}$~(fm$^{-1}$) &$\langle \rho\sigma\rangle_T^{{\rm NR}}$~(fm$^{-1}$) \\
\hline
$0.150$  &$-$ &$0.00700$&$0.00182$&$0.00244$&$0.00052$ &$0.00469$ &$0.01648$\\
      &$0.00435$ &$0.00801$&$0.00212$&$0.00268$&$0.00052$ &$0.00580$ &$0.01908$\\ \hline
$0.155$  &$-$ &$0.00948$&$0.00239$&$0.00341$&$0.00074$ &$0.00607$ &$0.02208$\\
      &$0.00520$ &$0.01101$&$0.00280$&$0.00375$&$0.00074$ &$0.00753$ &$0.02565$\\ \hline
$0.160$  &$-$ &$0.01267$&$0.00311$&$0.00467$&$0.00102$ &$0.00774$ &$0.02920$\\
      &$0.00622$ &$0.01478$&$0.00365$&$0.00516$&$0.00102$ &$0.00967$ &$0.03402$\\ \hline
$0.165$  &$-$ &$0.01672$&$0.00398$&$0.00631$&$0.00138$ &$0.00977$ &$0.03817$\\
      &$0.00737$ &$0.01959$&$0.00470$&$0.00670$&$0.00138$ &$0.01224$ &$0.04456$\\ \hline
$0.170$  &$-$ &$0.02183$&$0.00505$&$0.00842$&$0.00186$ &$0.01219$ &$0.04935$\\
      &$0.00868$ &$0.02566$&$0.00597$ &$0.00934$&$0.00186$ &$0.01533$ &$0.05769$\\ \hline
$0.175$  &$-$ &$0.02821$&$0.00633$&$0.01109$&$0.00247$ &$0.01506$ &$0.06316$\\
      &$0.01010$ &$0.03326$&$0.00751$&$0.01234$&$0.00247$ &$0.01904$ &$0.07398$\\ \hline
$0.180$  &$-$ &$0.03610$&$0.00786$&$0.01445$&$0.00324$ &$0.01845$  &$0.08010$\\
      &$0.01175$ &$0.04270$&$0.00935$&$0.01612$&$0.00324$ &$0.02341$ &$0.09400$\\ \hline
%$0.185$  &$-$ &$0.04578$&$0.00967$&$0.01863$&$0.00312$ &$0.02241$  &$-$\\
%      &$0.01356$ &$0.05432$&$0.01154$&$0.02085$&$0.00311$ &$0.02855$ &$0.11838$\\    %\hline
%$0.190$  &$-$ &$0.05756$&$0.01180$&$0.02379$&$0.00401$ &$0.02702$  &$-$\\
%      &$0.01560$ &$0.06853$&$0.01412$&$0.02667$&$0.00400$ &$0.03458$ &$0.14794$\\    %\hline
%$0.195$  &$-$ &$0.07181$&$0.01428$&$0.03010$&$0.00511$ &$0.03234$  &$-$\\
%      &$0.01782$ &$0.08577$&$0.01715$&$0.03389$&$0.00510$ &$0.04159$ &$0.18350$\\    %\hline
%$0.200$  &$-$ &$0.08891$&$0.01717$&$0.03778$&$0.00646$ &$0.03847$  &$-$\\
%      &$0.01990$ &$0.10646$&$0.02067$&$0.04259$&$0.00644$ &$0.04961$ &$0.22579$\\ 
\hline
\end{tabular}
\caption{Inverse absorption lengths as defined in Eq.~(\ref{length}) for all the particles in the gas.
For each temperature we show the results obtained for fixed $D$-mesons masses (upper entry of each cell) 
and for decreasing $D$-mesons masses as computed in~\cite{Fuchs:2004fh} (lower entry of each cell).
Since the $\phi$ decays only into $D_s\bar{D}_s$ and we assume that $D_s$ mesons do not change their masses and widths
inside a hadron medium, the upper and lower entry of each cell are equal.
As for the resonant contribution due to $X_2$ (first column) we do not report the results with fixed $D$-mesons masses, since they are negligible 
with respect to the non-resonant ones (NR). We do not consider temperatures higher than the value we use for the Hagedorn temperature $T_H\sim 177$~MeV.}
\label{tab:absorption}
\end{table}
\end{center}

We now take into account the recent RHIC data on the so called nuclear modification factor $R^{^{J/\psi}}_{_{A+A}}$, 
reported in~\cite{Adare:2006ns} as a function of the number of participants in the collision.
The quantity $R^{^{J/\psi}}_{_{A+A}}$ measures the ratio of the $J/\psi$ yield in $A+A$ and $pp$ collisions scaled by the number of nucleon-nucleon collisions. 
We will consider only Au-Au collisions at RHIC, due to their higher statistical significance.
We will also reconsider the old data on Pb-Pb collisions from NA50~\cite{Alessandro:2004ap} to show how the picture has changed in the last years.

Refs.~\cite{Thews:2005vj,Abreu:2007kv} have also considered the possibility that a recombination mechanism could compensate the $J/\psi$ suppression due to QGP,
making the drop in RHIC data less evident with respect to NA50, where this mechanism is expected to be weaker due to the much smaller energies involved.
However in~\cite{Capella:2007jv} it was shown that the recombination effects are of the same order of magnitude as the experimental uncertainties and thus they can be safely neglected.

\begin{figure}[!h]
\includegraphics[width=8truecm]{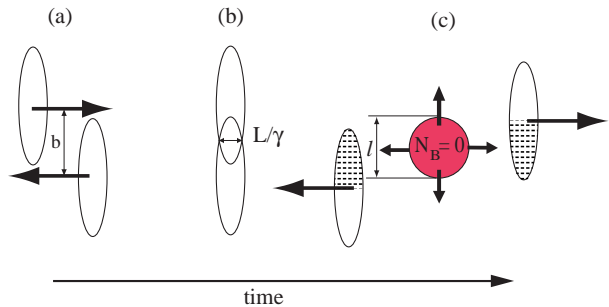}
\caption{Geometry of the collision between two identical heavy nuclei with impact parameter $b$. After the two nuclei have traversed each other, a thermalized gas of lighter resonances is formed.}
\label{fig:coll}
\end{figure}

The geometry of the heavy ion collision is shown schematically in Fig.~\ref{fig:coll}, which depicts the time-evolution in the center of mass frame.
The impact parameter, $b$, is defined as the transverse distance of the centers of the two nuclei. We consider the $J/\psi$ to be created with FeynmanÕs $x\simeq 0$, 
during the overlap of the two nuclei. These particles have to overcome absorption from the column density of nucleons of extension $L$. 
In the center of mass frame the length of the column is $L/\gamma$. In the same frame, the density of nucleons is $\rho_{{\rm nucl}}\gamma$, 
so that the absorption factor is Lorentz invariant and given by $\exp\left(-\rho_{{\rm nucl}}\sigma_{{\rm nucl}}L\right)$: see~\cite{Abreu:1999qw}. 
The nuclear absorption cross-section, $\sigma_{{\rm nucl}}$, has been determined in~\cite{Tram:2008zz} from the behavior of the cross-section for 
$p+A\to J/\psi+{\rm Anything}$ and ${\rm d}+{\rm Au}\to J/\psi + {\rm Anything}$
\be
\sigma_{{\rm nucl}}^{^{\rm RHIC}} = (3.5 \pm 0.2)~{\rm mb}.
\ee
As for NA50 one learns from~\cite{Cortese:2003iz} that
\be
\sigma_{{\rm nucl}}^{^{\rm NA50}} = (4.3 \pm 0.2)~{\rm mb}.
\ee
For the density of ordinary nuclear matter we take $\rho_{{\rm nucl}} = 0.17~{\rm fm}^{-3}$~\cite{Letessier:2002gp}.

In Fig.~\ref{fig:coll}(c) we show the hadron fireball produced by the central collisions of the interacting nucleons~\cite{Bjorken:1982qr} (the {\it comoving particles} $\pi,\rho,\omega, ...$). 
The fireball has a transverse dimension, $l$, approximately  equal to the length of the overlapping region
\be
\label{eq:elle}
{\it l}=2R-b,
\ee
where $R$ is the nuclear radius. 
The attenuation due to the interactions with the hadrons in the fireball is related to the average length that a $J/\psi$ has to traverse before leaving it.
The RHIC data in~\cite{Adare:2006ns} are taken in two different rapidity regions: a forward rapidity region $1.2<|y|<2.2$ and a mid rapidity region $|y|<0.35$. 
We take, for simplicity, a spherical fireball and we simulate the production of a particle at some point inside the sphere
and with a given direction of the velocity. 
Assuming a uniform linear motion inside the fireball, the distance $d$ given the starting point $\bm{r}$ and the direction of the velocity $\hat{\bm{v}}$,
can be written in implicit form as
\be
|\bm{r}+d\hat{\bm{v}}|=\frac{l}{2}.
\ee
The point on the spherical surface where the particle emerges from the fireball is thus $\bm{r}'=\bm{r}+d\hat{\bm{v}}$,
from which one can compute the rapidity of the $J/\psi$ observed, $y\simeq\eta= -\ln(\tan(\theta/2))$,
with $\theta$ the  polar angle associated to $\bm{r}'$.
%\footnote{We are assuming that the rapidity $y=\frac{1}{2}\ln((E+p_z)/(E-p_z))$ of the $J/\psi$ can be approximated by its pseudorapidity $\eta=-\ln(\tan(\theta/2))$. 
%This approximation is valid as far as $|\bm{p}_\psi|>>m_\psi$.}
To obtain the average distance one needs to integrate over the two angles which identify the direction of $\hat{\bm{v}}$ and over $\bm{r}$,
taking into account the constraint on the polar angle of the emersion point given by the experimental bounds on the rapidity ($y_0<|y|<y_1$ implies $\theta_0<\theta<\theta_1$).
To make our simple simulation more realistic we take into account that the distribution of the directions of the velocity is not uniform, 
but can be approximated by  $f(\hat{\bm{v}}_{_T})\propto 1/p_{_T}$, where $p_{_T}$ is the transverse momentum with respect to the beam axis.
Finally one has
\be
\bar{d} =\frac{\int d\bm{r}\int d\hat{\bm{v}}\;f(\hat{\bm{v}}_{_T})d(\bm{r},\hat{\bm{v}})\;T(\theta)}{\int d\bm{r}\int d\hat{\bm{v}}\;f(\hat{\bm{v}}_{_T})\;T(\theta)},
\ee
with 
\be
T(\theta)=\begin{cases}&1\;\;\;\;\theta_0<\theta<\theta_1\\ & 0 \;\;\;\;{\rm elsewhere}\end{cases}.
\ee
The result of this computation is $\bar{d}_{_{\rm fwd}}=0.4\,l$ and $\bar{d}_{_{\rm mid}}=0.3\,l$ in the forward and mid rapidity region respectively.  
Thus, the attenuation factor due to absorption by the comoving particles is
\be
\label{eq:abscomov}
{\mathcal A}^{^{\rm fwd(mid)}}_{\pi,\rho,\omega,...}\propto {\rm exp}\left[-\Sigma_i \langle \rho_i\sigma_i\rangle_{T(l)} \bar{d}_{_{\rm fwd(mid)}}\right],
\ee

\noindent the subscript $i$ labels the species of hadrons making up the fireball, ${\rho}_i$ the number density of the effective (i.e. above threshold)
particles and ${\sigma}_{i}$ the corresponding $J/\psi$ dissociation cross-section. Brackets indicate an average over the energy distribution 
in the fireball. This thermal average is computed at a certain temperature $T(l)$, which is given by the centrality of the collision, as we shall explain in detail in the next subsection.

The NA50 measurements on Pb-Pb collisions were inclusive. Hence one needs to integrate the distance $d(\bm{r},\hat{\bm{v}})$ over the whole range for the polar angle,
obtaining $\bar{d}=3/8\;l$, as was done in the previous analysis contained in~\cite{Maiani:2004qj}.

\begin{figure}[!h]
\begin{center}
\begin{minipage}[t]{7truecm} % A minipage that covers half the page
\centering
\includegraphics[width=7truecm]{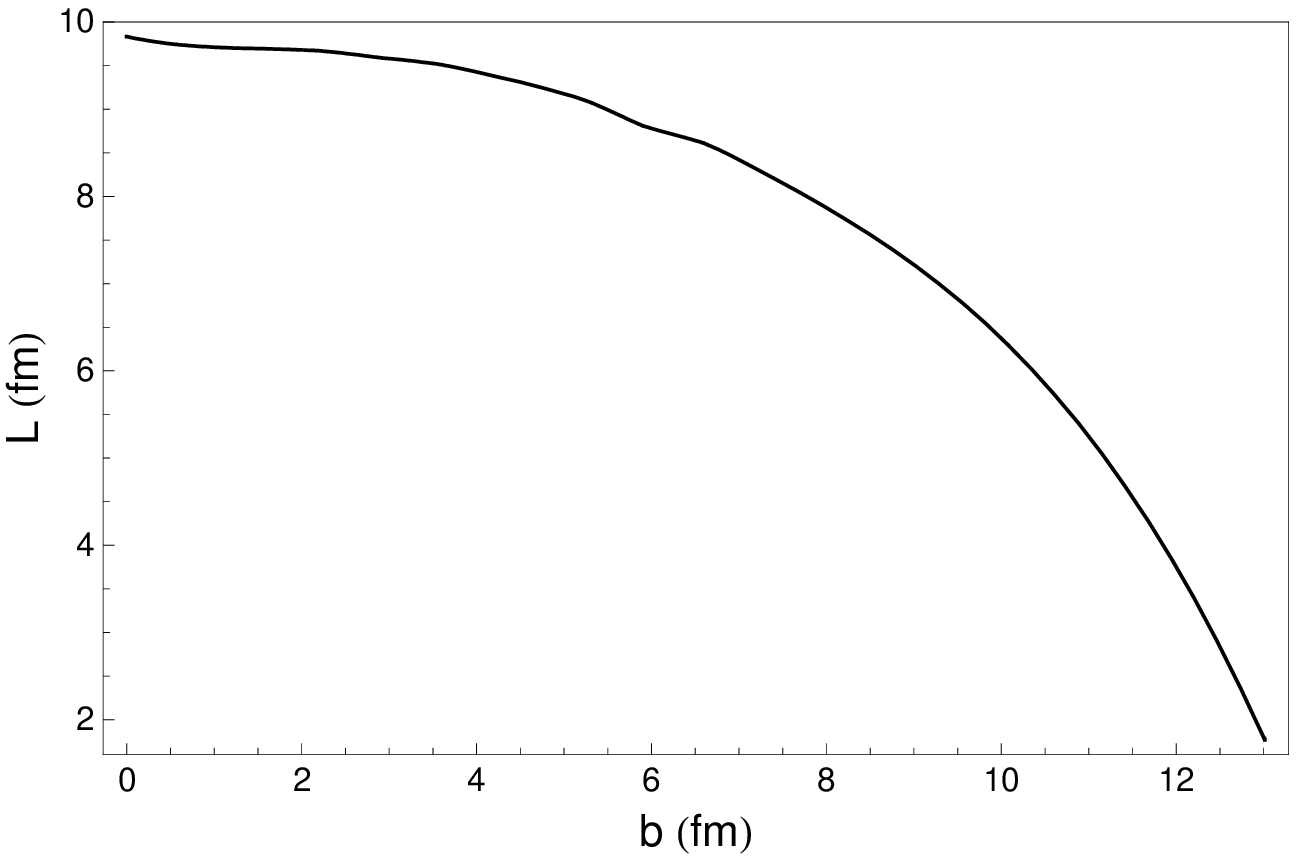}
\caption{Mean length of the path that a $J/\psi$ produced during a Pb-Pb collisions at NA50 must travel in nuclear matter as a function of the impact parameter $b$~\cite{Abreu:1997jh,Abreu:1997ji}.}
\label{fig:elledib}
\end{minipage}
\hspace{1truecm} % To get a little bit of space between the figures
\begin{minipage}[t]{7truecm}
\centering
\includegraphics[width=7truecm]{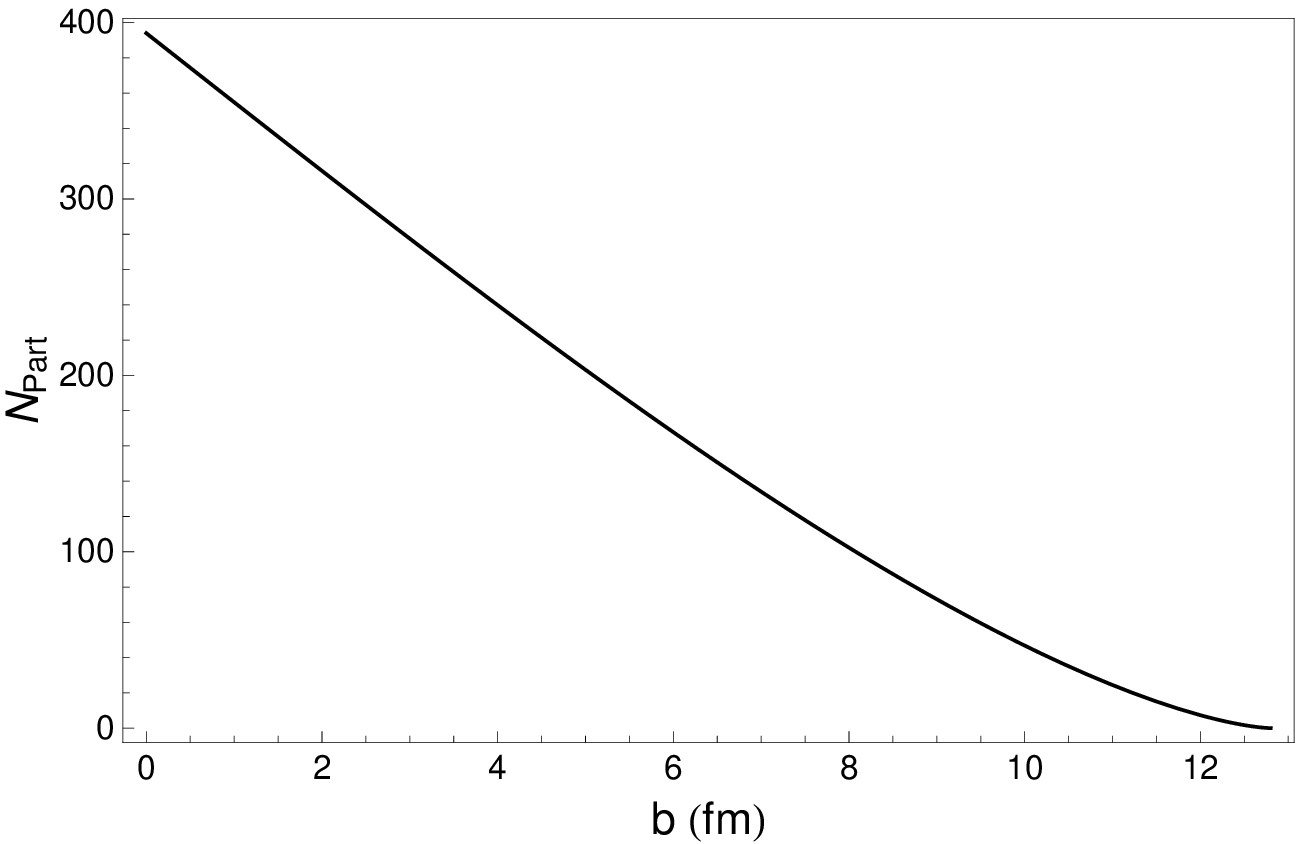}
\caption{Average number of participant nucleons in a Au-Au collision as a function of the impact parameter $b$ computed using Eq.~(\ref{npart}).}
\label{fig:npart}
\end{minipage}
\end{center}
\end{figure}

As noted before, we can compute the nuclear absorption length, $L$, as a function of $b$ using NA50 data~\cite{Abreu:1997jh,Abreu:1997ji} for Pb-Pb collisions. 
We report this function in Fig.~\ref{fig:elledib}.
Exploiting Eq.~(\ref{eq:elle}), one can obtain $L$ as a function of ${\it l}$.
\noindent We can reasonably suppose that the same function $L(b)$ can be used in the analysis of Au-Au collisions at RHIC,
since Au and Pb have approximately the same radius~($R_{_{\rm Pb}}=7.1~{\rm fm}$ and $R_{_{\rm Au}}=7.0~{\rm fm}$).

Putting it all together, we write the attenuation of the $J/\psi$, due to both comovers and nuclear effects, as a function of $\it l$ according to

\be
\label{eq:attot}
\mathcal{A}^{^{\rm fwd(mid)}}(l)=C_0+C\times {\rm exp}[-\rho_{\rm nucl}\sigma_{\rm nucl} L(l)]\times {\rm exp}\left[-\Sigma_i \langle \rho_i\sigma_i\rangle_{T(l)} \bar{d}_{_{\rm fwd(mid)}}\right],
\ee

\noindent where $C$ is an appropriate normalization constant and $C_0$ is an offset. To fit NA50 data we substitute $\bar{d}_{_{{\rm fwd(mid)}}}$ with $3/8\; l$.

To obtain the experimental data~\cite{Adare:2006ns} as a function of $l$ we derive the number of nucleons participating in a collision with impact parameter $b$ from geometrical considerations
\be
\label{npart}
N_{{\rm part}}=4\left[\int_{0}^{\sqrt{R^2-b^2/4}}d\!x\;\sqrt{R^2-x^2}-\frac{b}{2}\sqrt{R^2-b^2/4}\right]\frac{2A}{\pi R^2},
\ee
\noindent where $A$ is the atomic mass number and $R$ the nucleus radius, see Fig.~\ref{fig:npart}. 
Very similar results are obtained using a Glauber Model to relate the number of participants to the impact parameter~\cite{Alessandro:2004ap}.

\subsection{Hagedorn Gas}
\label{subsec:hagedorn}

Here we wish to determine the function $T(l)$ to be used in Eq.~(\ref{eq:abscomov}).
We will describe the fireball as a Hagedorn gas of resonances. 
The partition function of a Hagedorn gas in the classical Boltzmann limit ($E>> \kappa_B T$) can be written as~\cite{Letessier:2002gp}
\be
\label{partfunc}
\ln(Z^{cl}_{_H})=\left(\frac{T}{2\pi}\right)^{3/2}\int\;dm\;\rho(m)m^{3/2}e^{-m/T}.
\ee
$\rho(m)$ is the mass spectrum of hadronic states, which has the empirical shape
\be
\rho(m)=\frac{c}{\left(m_0^2+m^2\right)^{3/2}}e^{m/T_{_H}},
\ee
\noindent with $m_0= 0.96~{\rm GeV}$, $c=2.12~{\rm GeV}^2$ and $T_{_H}=177~{\rm MeV}$~\cite{Maiani:2004qj}. 
$T_H$ is known as the Hagedorn temperature. For a recent determination see~\cite{Cleymans:2011fx}.
As soon as $T\geq T_{_H}$ the integral in Eq.~(\ref{partfunc}) diverges, hence this thermodynamical description is valid up to $T\leq T_{_H}$. 
Above the Hagedorn temperature the system undergoes a phase transition, which can be interpreted as the transition from hadronic matter to QGP.

\begin{figure}[!h]
\begin{center}
\begin{minipage}[t]{7truecm} % A minipage that covers half the page
\centering
\includegraphics[width=8truecm]{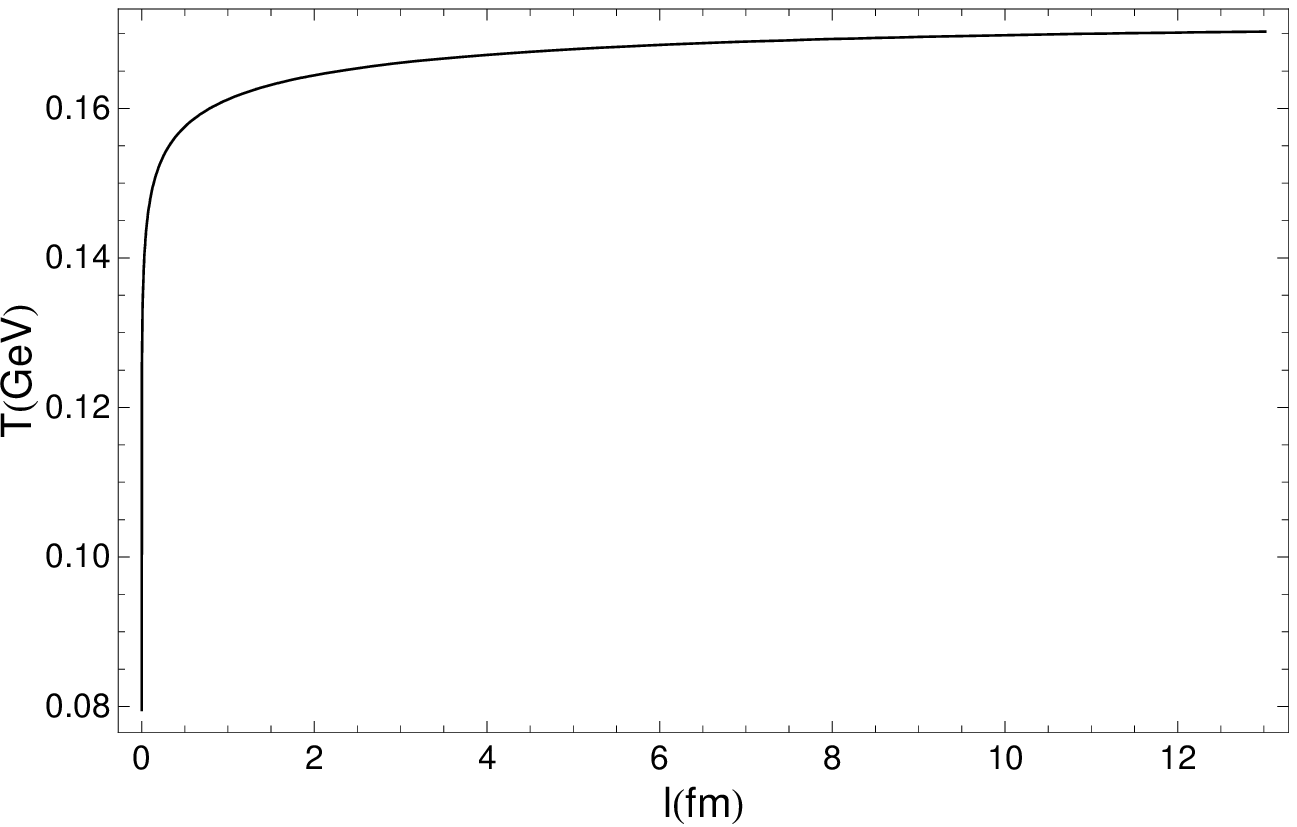}
\end{minipage}
\hspace{1truecm} % To get a little bit of space between the figures
\begin{minipage}[t]{7truecm}
\centering
\includegraphics[width=8truecm]{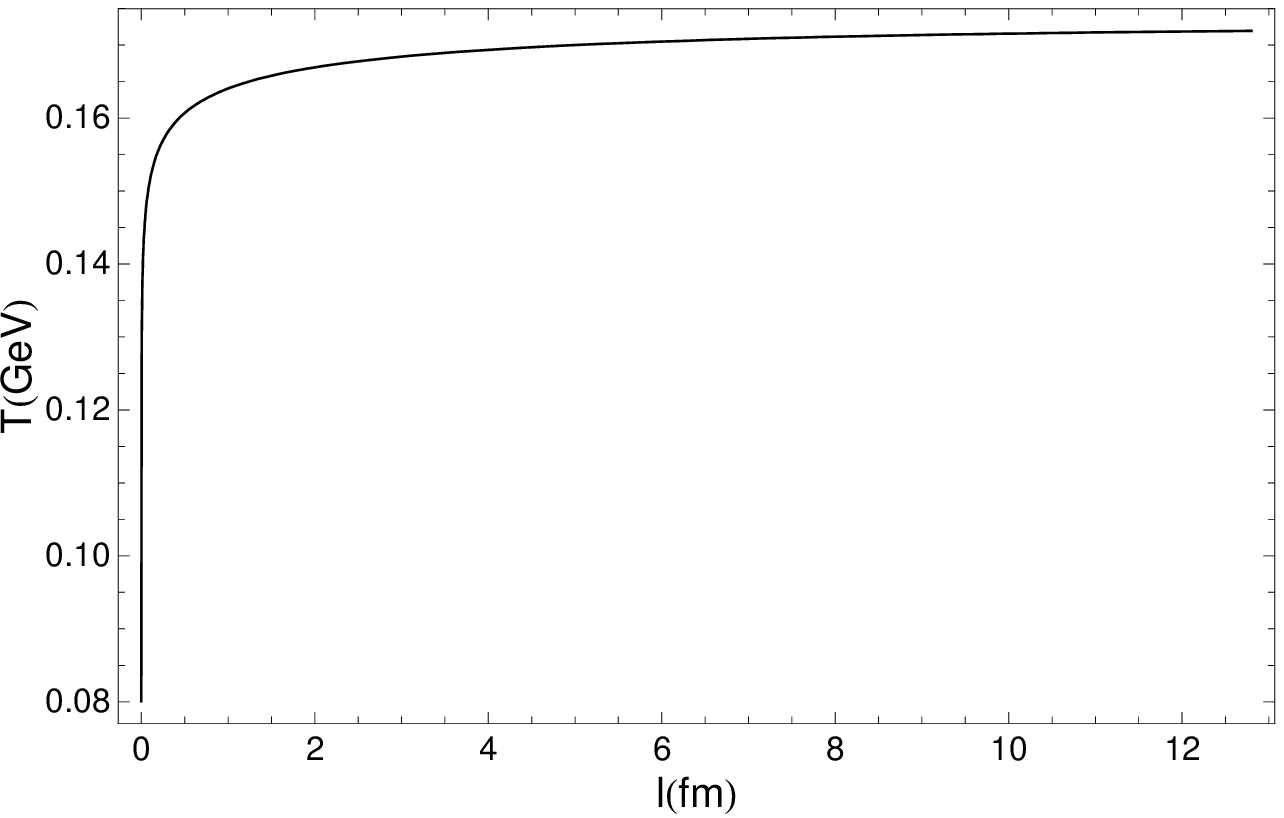}
\end{minipage}
\caption{Temperature of the Hagedorn gas formed after a Pb-Pb collision at NA50 (left panel) or a Au-Au collision at RHIC (right panel) as a function of $l=2R-b$, $b$ being the impact parameter.}
\label{fig:Tdib}
\end{center}
\end{figure}

\begin{figure}[!h]	
\includegraphics[width=9truecm]{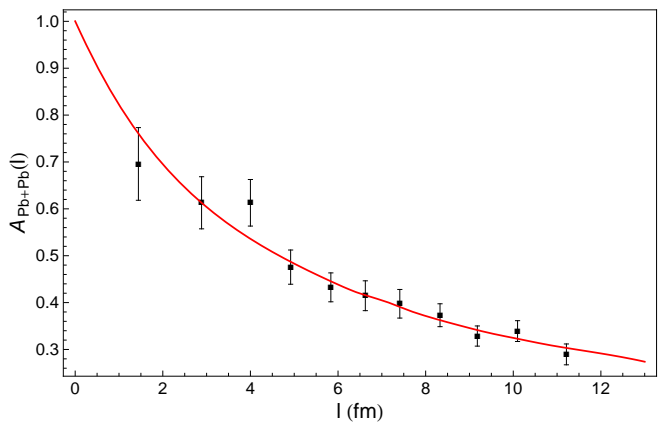}
\caption{Attenuation function for the $J/\psi$ yield in Pb-Pb collisions as measured at NA50 (Squares) and as predicted by the hadron gas description (Red Line). The best fit is obtained for $C_0=-0.2$ and $C=1.4$ giving a $\chi^2/{\rm DOF}=4.9/9$. In absence of the resonant contribution from $X_2$ and of the in-medium effects on the $D$ mesons we obtain $\chi^2/{\rm DOF}=5.1/9$.}
\label{fig:fitPb}
\end{figure}

\begin{figure}[!h]
\begin{center}
\begin{minipage}[t]{7truecm} % A minipage that covers half the page
\centering
\includegraphics[width=8truecm]{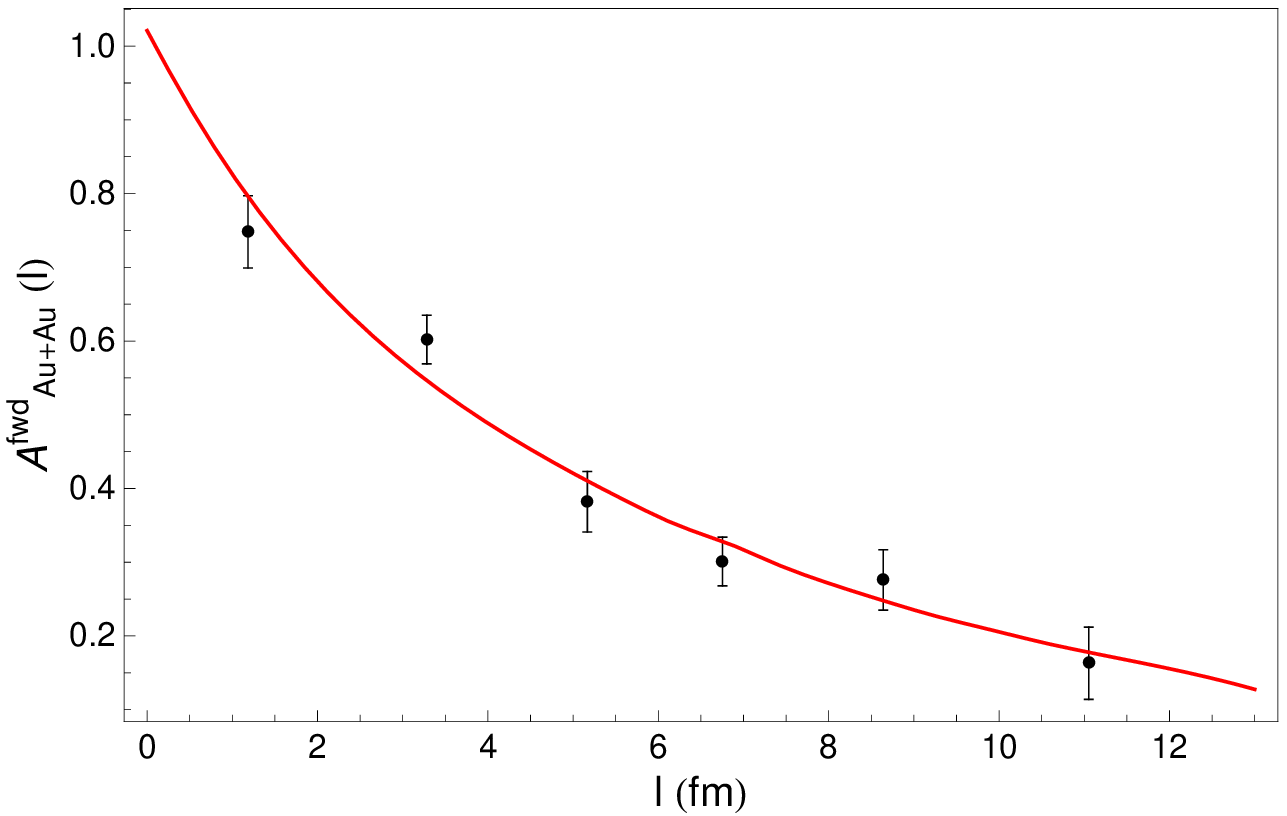}
\caption{Attenuation function for the $J/\psi$ yield in Au-Au collisions in the forward rapidity region $1.2<|y|<2.2$ as measured at RHIC (Disks) and as predicted by the hadron gas description (Red Line).
The best fit is obtained for $C_0=-0.6$ and $C=1.8$ giving a $\chi^2/{\rm DOF}=5.5/4$. In absence of the resonant contribution from $X_2$ and of the in-medium effects on the $D$ mesons we obtain $\chi^2/{\rm DOF}=6/4$. 
In this rapidity region the $J/\psi$ is reconstructed in the $\mu^+\mu^-$ mode.}
\label{fig:fitAufwd}
\end{minipage}
\hspace{1truecm} % To get a little bit of space between the figures
\begin{minipage}[t]{7truecm}
\centering
\includegraphics[width=8truecm]{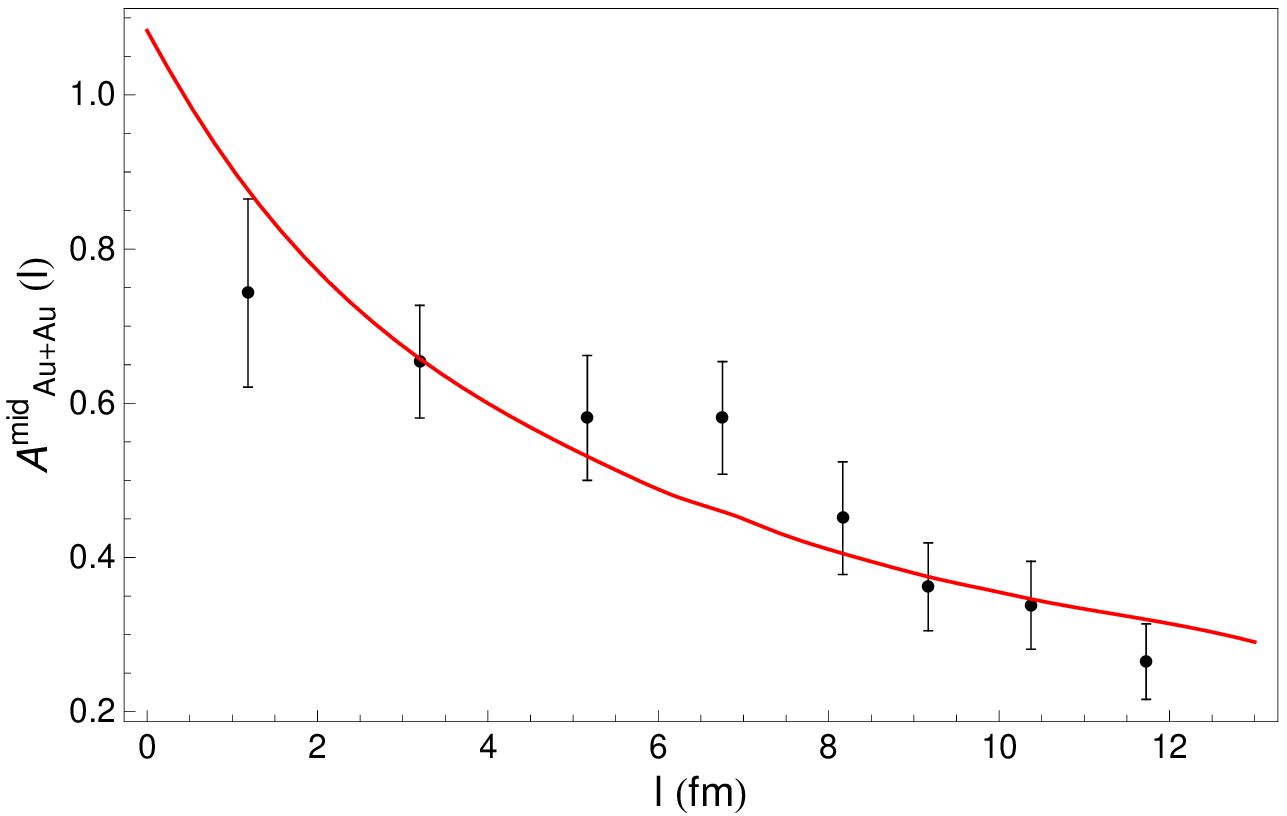}
\caption{Attenuation function for the $J/\psi$ yield in Au-Au collisions in the mid rapidity region $|y|<0.35$ as measured at RHIC (Disks) and as predicted by the hadron gas description (Red Line).
The best fit is obtained for $C_0=-0.4$ and $C=1.7$ giving a $\chi^2/{\rm DOF}=6/6$. In absence of the resonant contribution from $X_2$ and of the in-medium effects on the $D$ mesons we obtain $\chi^2/{\rm DOF}=6.5/6$.
In this rapidity region the $J/\psi$ is reconstructed in the $e^+e^-$ mode.}
\label{fig:fitAumid}
\end{minipage}
\end{center}
\end{figure}

From the partition function of Eq.(\ref{partfunc}) one can easily obtain the the energy density $\epsilon(T)$
\be
\label{energia}
\epsilon(T)=-\frac{\partial}{\partial \beta}\ln(Z^{cl}_{_H})=\left(\frac{T}{2\pi}\right)^{3/2}\int\;dm\;\frac{c}{\left(m_0^2+m^2\right)^{3/2}}m^{5/2}\left(1+\frac{3}{2}\frac{T}{m}\right)e^{m(1/T_{_H}-1/T)}.
\ee
On the other hand the energy density released in a collision is proportional to the factor
\be
\frac{\rho_{{\rm nucl}}V(b)}{S(b)}=\frac{A}{S}g(b/R),
\ee
\noindent where
\be
g(b/R)=\frac{\pi}{2}\frac{\left(1-b/2R\right)^2 \left(b/4R+1\right)}{\arccos\left(b/2R\right)-(b/2R) \sqrt{1-b^2/4R^2}}.
\ee
Therefore a simple estimate of the ratio of the energy density for two different values of $b$ is given by
\be
\frac{\epsilon(b)}{\epsilon(b_0)}=\frac{g(b/R)}{g(b_0/R)}\Rightarrow \epsilon(b)=\frac{\epsilon(b_0)}{g(b_0/R)}g(b/R).
\ee
Using the Bjorken relation~\cite{Bjorken:1982qr} one can estimate the energy density released in a collision with impact parameter $b$
by measuring the transverse energy per unit rapidity
\be
\epsilon_{_{Bj}}=\frac{dE_{_T}}{dy}\frac{1}{ \tau_0\pi r^2},
\ee
\noindent where $\tau_0$, the formation time, is usually taken as $1~{\rm fm}$, and $\pi r^2$ is the effective area of the collision.
In~\cite{Adler:2007fj} the PHENIX collaboration finds that in Au-Au collisions with $90$ participants, which corresponds to $b_0\simeq 9~{\rm fm}$ (see Eq.~(\ref{npart})),
the energy density amounts to $2.4~{\rm GeV}/{\rm fm}^3$, thus
\be
\epsilon^{^{{\rm RHIC}}}(b_0=9~{\rm fm})=2.4~{\rm GeV}/{\rm fm}^3.
\ee
As for the NA50 data on Pb-Pb collisions, we take from~\cite{Abreu:2002fx}
\be
\epsilon^{^{{\rm NA50}}}(b_0=9.2~{\rm fm})=1.9~{\rm GeV}/{\rm fm}^3.
\ee
Using these values we obtain the energy density as a function of the impact parameter $\epsilon(b)$. 
On the other hand we know the relation between energy density and temperature $\epsilon(T)$ from Eq.~({\ref{energia}}) and thus we 
can deduce $T(b)$ and in turn $T(l)$. We show $T(l)$ for Pb-Pb collisions at NA50 and Au-Au collisions at RHIC in Fig.~\ref{fig:Tdib}.
It is evident that over a wide range of $l$ the temperature is almost constant and below the Hagedorn temperature.
Now we have all the ingredients to perform a best fit of the experimental data using the attenuation function defined in Eq.~(\ref{eq:attot}).
We show the agreement between experimental data and theoretical prediction in Fig.~\ref{fig:fitPb} for NA50 data, and in Fig.~\ref{fig:fitAufwd} and Fig.~\ref{fig:fitAumid} for RHIC data.

\section{Discussion and Conclusions}
We have determined from available data the strong coupling constants of the $X(3872)$ under the hypothesis that it is a $1^{++} (X_1)$ or $2^{-+}(X_2)$ resonance. The results we find may be confronted with theoretical calculations making assumptions on the structure of the $X$: molecule, tetraquark, standard charmonium. 

We use the coupling strengths we find to explore the potential role of $X_{1,2}$ in $J/\psi$ absorption processes like $J/\psi\; (\rho,\omega)\to X_{1,2}\to D^0\bar D^{0*}$. Such processes might occur in a hot resonance gas produced in  heavy-ion collisions. 
Regardless of the detail mechanism by which the resonance gas is formed, processes as the ones mentioned above are mimicking the in-plasma $J/\psi$ suppression hypothetically due to the Debye screening of the $c\bar c$ confining potential, as first discussed in~\cite{Matsui:1986dk}. Therefore this is a background to the Debye $J/\psi$ suppression signal. How far  can we go with a hadron gas picture  in fitting data on $J/\psi$ suppression at RHIC? A limitation to the hadron gas description might come from the excessively high temperature needed for the gas to account for the observed $J/\psi$ suppression effect. This was discussed in~\cite{Maiani:2004py,Maiani:2004qj}:  a hadron gas description fails above the critical Hagedorn temperature, the highest temperature for hadron matter.   

The analysis in~\cite{Maiani:2004py,Maiani:2004qj} was based on NA50 data on Pb-Pb collisions,  see Fig.~\ref{fig:fitPb}, where  at a centrality of about 4 fm in units of $l=2R-b$, $b$ being the impact parameter,  a drop was observed (actually a one sigma effect) in the $J/\psi$ yield in going from the three leftmost points  towards higher centralities. In the low $l$ region  $(l\leq 4~{\rm fm})$ the authors of~\cite{Maiani:2004py,Maiani:2004qj} used also data on S-U collisions and the approach was to perform a best fit in that region (where the hadron gas picture is more reliable, the energy density being smaller) with an exponential attenuation function at some temperature $T$. 
An unreasonably large $T$ was then needed to fit data also at $l\geq 4$~fm. Using an Hagedorn  gas model the fit was simply very poor. 
Moreover Ref.~\cite{Becattini:2005yj} pointed out a correlation in the $l$ dependence of the $J/\psi$ suppression and the enhancement of strange particle production
observed in NA50 data.

Here we take a different approach. We note that the drop at $l=4$~fm observed by NA50 
(Fig.~\ref{fig:fitPb}), is much less evident in the recent Au-Au RHIC data in Fig.~\ref{fig:fitAufwd} and absent in Fig.~\ref{fig:fitAumid}.
Therefore we fit the whole data set (and not only the $l\leq 4$~fm region) with an attenuation function computed in a Hagedorn gas having a limiting temperature $T_H=177$~MeV.
As stated in Section~\ref{subsec:data} we are neglecting  possible charm recombination effects.

Actually we find a very good fit to data just using the attenuation functions computed in~\cite{Maiani:2004qj}. 
This is so because the nuclear part of the attenuation function in Au-Au collisions at RHIC is expected to be almost the same 
as that in Pb-Pb collisions at NA50, because Pb and Au nuclei are very similar in size and the $J/\psi$ 
nuclear absorption cross sections turns out to be very similar at RHIC and NA50.
Moreover to define the dependence of the temperature on centrality, we use the energy density produced in the RHIC collisions according to the Bjorken description, and we found it to be almost equal to the one
computed for NA50. 

Including some hypothetical in-medium effects on mass decreasing and broadening of open charm mesons~\cite{Faessler:2002qb,Fuchs:2004fh,Tolos:2007qr,Tolos:2007vh,Gottfried:1992bz,Hayashigaki:2000es,Kumar:2009xc,Kumar:2010gb,Kumar:2010nz} and a resonant contribution from the
$J/\psi\;(\rho,\omega) \to X_2 \to D^0 \bar{D}^{0*}$ channel (about the $15\%$ of the non-resonant one)
we get slightly larger inverse absorption lengths as shown in Table~\ref{tab:absorption} which altogether slightly improve the fit to data, decreasing the $\chi^2/{\rm DOF}$ from $6/4 $ to $5.5/4$; see Fig.~\ref{fig:fitAufwd} and~\ref{fig:fitAumid}. 
In the calculation of $J/\psi$ absorption we assume that $X=X_2$ has more likely a $2^{-+}$ charmonium interpretation, 
whereas $X=X_1$ has a $1^{++}$ molecule assignment if only because we have no clues on how a finite temperature hadron medium would alter mass and width of a tetraquark particle.
In this respect a charmonium $X_2$ gets a much larger width because its mass is not modified by the medium, while the masses of its decay products $D^0$ and $D^{0*}$ are.
The larger width of $X_2$ is in turn responsible for the most effective $J/\psi\;(\rho,\omega)\to X_2$ conversion which has a very low rate for a narrow $X_1$.
The $X_1$ is expected to stay narrow even in medium because: $i)$ its mass lowers as the sum of $D^0$ and $D^{0*}$ masses, $ii)$ $X_1\to D^0\bar{D}^{0*}$ is an {\it S}~-~wave decay.

The conclusion which can be drawn from this analysis is that, given the hypotheses we use, data on $J/\psi$ yield in heavy-ion collisions are likely the less compelling ones in the search of a deconfined quark-gluon state of matter because they are affected by a large hadronic background in the sense explained above.

In order to have a clearer picture, it would be very important to have RHIC data on $J/\psi$ suppression for a larger number of impact parameter bins, in particular in the intermediate centrality range.
It is moreover our aim to underscore that $X,Y,Z$ particles being discovered in the last few years might have impact on a wide class of elementary processes: we find here that if the $X(3872)$ were confirmed to be a $2^{-+}$ state, under certain hypotheses on the behavior of open charm mesons in a hot hadron gas,  it would give a non-negligible contribution to the hadron $J/\psi$ dissociation mechanism.

\appendix

\section{$J/\psi\;(\rho,\omega)\to X(3872)\to  D^0\bar{D}^{0*}$ cross-section}
\label{app:xsect}
We give some details on the formulae used in the text.
The differential cross-section for $J/\psi\;\rho\to D^0\bar{D}^{0*}$ is
\be
d\sigma(J/\psi\;\rho\to D^0\bar{D}^{0*})=\frac{1}{9}\frac{1}{4\phi}(2\pi)^4\delta^{(4)}\left(p_{_D}+p_{_{D^*}}-p_\psi-p_\rho\right)\sum_{{\rm pol}}\left|\M_{J/\psi\;\rho\to D^0\bar{D}^{0*}}\right|^2\frac{d^3\!p_{_D}}{(2\pi)^32\omega_{_D}}\frac{d^3\!p_{_D}}{(2\pi)^32\omega_{_{D^*}}},
\ee
\noindent with the flux $\phi$ defined by
\be
\phi=\sqrt{(p_\psi\cdot p_\rho)^2-m_\psi^2m_\rho^2}=\frac{1}{2}\sqrt{\lambda(s,m^2_\psi,m^2_\rho)}.
\ee
\noindent We use also
\be
\label{gammas}
\int
(2\pi)^4\delta\left(\omega_{_D}+\omega_{_{D^*}}-\sqrt{s}\right)\delta^{(3)}({\bf p}_{_D}+{\bf p}_{_{D^*}})\frac{d^3\!p_{_D}}{(2\pi)^32\omega_{_D}}\frac{d^3\!p_{_D}}{(2\pi)^32\omega_{_{D^*}}}=\frac{1}{16\pi}\frac{\sqrt{\lambda(s,m^2_{_D},m^2_{_{D^*}})}}{s}\;d\cos\theta.
\ee
The above formulae leads to Eq.~(\ref{xsect}). Similarly for $\omega$.

\section{$X\to J/\psi\;\rho$}
\label{app:rho}
Here we report the formulae used for the computation of the width of $X\to J/\psi \;\rho$.
\be
d\Gamma(X\to J/\psi\;\pi^+\pi^-)=\frac{1}{2s_X+1}\frac{1}{2m_{_X}}\sum_{\substack{{\rm pol}}}|\langle J/\psi\;\pi^+\pi^-|X\rangle|^2\;d\Phi^{(3)},
\ee
where
\be
d\Phi^{(3)}=(2\pi)^4\delta^{(4)}(P-p_\psi-p_1-p_2)\frac{d^3p_\psi}{(2\pi)^32E_\psi}\frac{d^3p_1}{(2\pi)^32E_1}\frac{d^3p_2}{(2\pi)^32E_2}.
\ee
Using the narrow width approximation for the $\rho$ and the unstable particle propagator
\be
\sum_{\substack{{\rm pol}}}|\langle J/\psi\;\pi^+\pi^-|X\rangle|^2=\frac{1}{3}\sum_{\substack{{\rm pol}}}|\langle J/\psi\;\rho|X\rangle|^2\frac{1}{(s-m_\rho^2)^2+(m_\rho\Gamma_\rho)^2}\sum_{\substack{{\rm pol}}}|\langle \pi^+\pi^-|\rho\rangle|^2,
\ee
where $\sum_{\substack{{\rm pol}}}|\langle \pi^+\pi^-|\rho\rangle|^2=g^2_{\rho\pi}$, with $g_{\rho\pi}$ a constant number.
The phase space factor can be rewritten as
\be
\begin{split}
{\rm d}\bar{\Phi}^{(3)}&=(2\pi)^4\int d^4p_\rho \delta^{(4)}(P-p_\psi-p_\rho)\delta^{(4)}(p_\rho-p_1-p_2)\frac{d^3p_\psi}{(2\pi)^32E_\psi}\frac{d^3p_1}{(2\pi)^32E_1}\frac{d^3p_2}{(2\pi)^32E_2}\\
&=\frac{1}{2\pi}\int\;ds\;(2\pi)^4\delta^{(4)}(P-p_\psi-p_\rho)\frac{d^3p_\rho}{(2\pi)^32\sqrt{s+|\bm{p}_\rho|^2}}\frac{d^3p_\psi}{(2\pi)^32E_\psi}\\
&\;\;\;\;\;\;\;\; \times(2\pi)^4\delta^{(4)}(p_\rho-p_1-p_2)\frac{d^3p_1}{(2\pi)^32E_1}\frac{d^3p_2}{(2\pi)^32E_2}.
\end{split}
\ee
The notation $d\bar{\Phi}^{(3)}$ is to indicate that we have an intermediate $\rho$.
Now we observe that
\be
\int (2\pi)^4\delta^{(4)}(P-p_\psi-p_\rho)\frac{d^3p_\rho}{(2\pi)^32\sqrt{s+|\bm{p}_\rho|^2}}\frac{d^3p_\psi}{(2\pi)^32E_\psi}=\Phi^{(2)}(m_{_X},m_\psi,\sqrt{s})=\frac{1}{4\pi}\frac{p^*(m^2_X,m^2_\psi,s)}{m_{_X}}
\ee
\noindent and
\be
\int (2\pi)^4\delta^{(4)}(p_\rho-p_1-p_2)\frac{d^3p_1}{(2\pi)^32E_1}\frac{d^3p_2}{(2\pi)^32E_2}=\Phi^{(2)}(\sqrt{s},m_{\pi^+},m_{\pi^-})=\frac{1}{4\pi}\frac{p^*(s,m^2_{\pi^+},m^2_{\pi^-})}{\sqrt{s}}.
\ee
Thus it results that
\be
\bar{\Phi}^{(3)}(m_X,m_\psi,m_{\pi^+},m_{\pi^-})=\frac{1}{2\pi}\int\;ds\;\frac{1}{4\pi}\frac{p^*(m^2_X,m^2_\psi,s)}{m_{_X}}\frac{1}{4\pi}\frac{p^*(s,m^2_{\pi^+},m^2_{\pi^-})}{\sqrt{s}}.
\ee
The full decay width is then
{\small \be
\begin{split}
\Gamma(X\to J/\psi\;\pi^+\pi^-)&=\frac{1}{2s_{_X}+1}\frac{1}{2m_{_X}}\frac{1}{6\pi}\int\;ds\;\sum_{\substack{{\rm pol}}}|\langle J/\psi\;\rho(s)|X\rangle|^2\frac{1}{4\pi}\frac{p^*(m^2_X,m^2_\psi,s)}{m_{_X}}\frac{g^2_{\rho\pi}}{(s-m_\rho^2)^2+(m_\rho\Gamma_\rho)^2}\frac{1}{4\pi}\frac{p^*(s,m^2_{\pi^+},m^2_{\pi^-})}{\sqrt{s}}.
\end{split}
\ee
}
We can relate $g^2_{\rho\pi}$ to $\Gamma(\rho\to\pi\pi)$ by
\be
g^2_{\rho\pi}=6 m^2_\rho\Gamma(\rho\to\pi\pi)\frac{4\pi}{p^*(m_\rho^2,m^2_\pi,m^2_\pi)}
\ee
{\small \be
\label{gammacompletarho}
\Gamma(X\to J/\psi\;\pi^+\pi^-)=\frac{1}{2s_X+1}\frac{1}{8\pi m^2_X}\int\;ds\;\sum_{\substack{{\rm pol}}}|\langle J/\psi\;\rho(s)|X\rangle|^2p^*(m^2_X,m^2_\psi,s)\frac{1}{\pi}\frac{m_\rho\Gamma_\rho\; \mathcal{B}(\rho\to\pi\pi)}{(s-m_\rho^2)^2+(m_\rho\Gamma_\rho)^2}\frac{m_\rho}{\sqrt{s}}\frac{p^*(s,m^2_{\pi^+},m^2_{\pi^-})}{p^*(m_\rho^2,m^2_\pi,m^2_\pi)}.
\ee
}
In the limit of narrow width for the $\rho$
\be
{\rm lim}_{\substack{{\Gamma\to0}}}\frac{m\Gamma}{(s-m^2)^2+(m\Gamma)^2}=\pi\delta(s-m^2).
\ee

\noindent Eq.~(\ref{gammacompletarho}) is equal to the one we can obtain taking the average of $\sum_{\substack{{\rm pol}}}|\langle J/\psi\;\rho(s)|X\rangle|^2 p^*(m^2_X,m^2_\psi,s)$ over the Breit-Wigner distribution of the $\rho$ meson
\be
\Gamma(X\to J/\psi\;\pi^+\pi^-)=\frac{1}{2s_X+1}\frac{1}{8\pi m^2_X}\int\;ds\;\sum_{\substack{{\rm pol}}}|\langle J/\psi\;\rho(s)|X\rangle|^2 p^*(m^2_X,m^2_\psi,s)\frac{1}{\pi}\frac{m_\rho\Gamma_\rho\; \mathcal{B}(\rho\to\pi\pi)}{(s-m_\rho^2)^2+(m_\rho\Gamma_\rho)^2}.
\ee
In actual calculations we use $m_\rho\Gamma_\rho\to \frac{s}{m_\rho}\Gamma_\rho$. 
Indeed the propagator of an unstable particle $A$ with 4-momentum $p$ and which decays to two particles $B$ and $C$ can be written in the form ($p^2=s$)~\cite{DeWit}:
\be
\frac{1}{(p^2-m^2)^2+(\sqrt{p^2}\;\Gamma(p^2))^2},
\ee
where $\Gamma(p^2)=\Gamma(A(p^2)\to BC)$ is
\be
\begin{split}
\Gamma(A(p^2)\to BC)=\frac{g^2(p^2,m^2_B,m^2_C)}{16\pi(\sqrt{p^2})^3}\sqrt{\lambda(p^2,m_B^2,m_C^2)}.
\end{split}
\ee
Even if an unstable state cannot be properly put on the mass-shell, 
the mass of a narrow resonance is still well defined and for $p^2$ equal to its mass its decay width is
\be
\begin{split}
\Gamma(A(m^2)\to BC)=\frac{g^2(m_A^2,m^2_B,m^2_B)}{16\pi m_A^3}\sqrt{\lambda(m_A^2,m_B^2,m_C^2)}.
\end{split}
\ee
It is then straightforward to see that
\begin{equation}
\label{eq:relgamma}
\Gamma(p^2)=\frac{m_A^3}{(\sqrt{p^2})^3}\;\frac{g^2(p^2,m^2_B,m^2_C)}{g^2(m^2_A,m^2_B,m^2_C)}\;\frac{\sqrt{\lambda(p^2,m_B^2,m_C^2)}}{\sqrt{\lambda(m_A^2,m_B^2,m_C^2)}}\;\Gamma(m_A^2).
\end{equation}
The coupling constant $g$ has the dimension of a mass, to give the right dimension to the width ($[g]=M$).
Now let us consider two limits which are relevant to our analysis. 
The first limit is the one in which both the particles in the final state are massless, {\it i.e.}, much lighter than $A$, $m_B=m_C=0$.
In this case the only mass scale of the problem is $p^2$ or $m_A^2$, and thus the only possibility is that $g^2(p^2,0,0)=\alpha p^2$ and thus $g^{2}(m^2_A,0,0)=\alpha m_A^2$,
where $\alpha$ is some adimensional constant.
The relation~(\ref{eq:relgamma}) reduces to
\begin{equation}
\label{eq:cw}
\Gamma(p^2)=\frac{\sqrt{p^2}}{m_A}\Gamma(m_A^2).
\end{equation}
This relation can be used when the unstable propagating particle has a mass much larger than that of its decay products, as in the case of the $\rho$.
For an intermediate $D^*$ we need to consider the case in which only one of the produced particles is massless $m_C=0$.
Since $\sqrt{\lambda(p^2,m^2_B,0)}=(p^2-m^2_B)$ one obtains:
\begin{equation}
\Gamma(p^{2})=\frac{m_{A}}{\sqrt{p^{2}}}\frac{(p^{2}-m^{2}_{B})}{(m^{2}_{A}-m^{2}_{B})}\Gamma(m^{2}_{A}).
\end{equation}

\section{$X\to J/\psi\;\omega$}
\label{app:omega}

Here we report the basic formulae for the computation of the width of $X\to J/\psi \;\omega$.
\be
d\Gamma(X\to J/\psi\;\pi^+\pi^-\pi^0)=\frac{1}{2s_X+1}\frac{1}{2m_{_X}}\sum_{\substack{{\rm pol}}}|\langle J/\psi\;\pi^+\pi^-\pi^0|X\rangle|^2d\Phi^{(4)},
\ee
where $d\Phi^{(4)}$ is
\be
d\Phi^{(4)}=(2\pi)^4\delta^{(4)}(P-p_\psi-\sum^4_{i=1}p_i)\frac{d^3p_\psi}{(2\pi)^32E_\psi}\prod^4_{i=1}\frac{d^3p_i}{(2\pi)^32E_i}.
\ee
Using narrow width approximation
\be
\sum_{\substack{{\rm pol}}}|\langle J/\psi\;\pi^+\pi^-\pi^0|X\rangle|^2=\frac{1}{3}\sum_{\substack{{\rm pol}}}|\langle J/\psi\;\omega|X\rangle|^2\frac{1}{(s-m_\omega^2)^2+(m_\omega\Gamma_\omega)^2}\sum_{\substack{{\rm pol}}}|\langle \pi^+\pi^-\pi^0|\omega\rangle|^2
\ee
and we further assume that $\sum_{\substack{{\rm pol}}}|\langle \pi^+\pi^-\pi^0|\rho\rangle|^2=g^2_{\omega\pi}$, with $g_{\omega\pi}$ a constant number.
The phase space factor can be rewritten as
\be
\begin{split}
d\bar{\Phi}^{(4)}&=(2\pi)^4\int d^4p_\omega \delta^{(4)}(P-p_\psi-p_\omega)\delta^{(4)}(p_\omega-\sum_{i}p_i)\frac{d^3p_\psi}{(2\pi)^32E_\psi}\prod_{i}\frac{d^3p_i}{(2\pi)^32E_i}\\
&=\frac{1}{2\pi}\int\;ds\;(2\pi)^4\delta^{(4)}(P-p_\psi-p_\omega)\frac{d^3p_\omega}{(2\pi)^32\sqrt{s+|\bm{p}_\omega|^2}}\frac{d^3p_\psi}{(2\pi)^32E_\psi}\\
&\;\;\;\;\;\;\; \times(2\pi)^4\delta^{(4)}(p_\omega-\sum_ip_i)\prod_i\frac{d^3p_i}{(2\pi)^32E_i},
\end{split}
\ee
where the notation $d\bar{\Phi}^{(4)}$ is to indicate that we have an intermediate $\omega$. Now
\be
\int (2\pi)^4\delta^{(4)}(P-p_\psi-p_\omega)\frac{d^3p_\omega}{(2\pi)^32\sqrt{s+|\bm{p}_\omega|^2}}\frac{d^3p_\psi}{(2\pi)^32E_\psi}=\frac{1}{4\pi}\frac{p^*(m^2_X,m^2_\psi,s)}{m_{_X}}
\ee
\noindent and
\be
\int (2\pi)^4\delta^{(4)}(p_\omega-\sum_ip_i)\prod_i\frac{d^3p_i}{(2\pi)^32E_i}=\Phi^{(3)}(\sqrt{s},m_{\pi^+},m_{\pi^-},m_{\pi^0}).
\ee
The expression for the three body phase space is the following
\be
\label{eq:ph3}
\Phi^{(3)}(\sqrt{s},m_1,m_2,m_3) = \frac{1}{32\pi^3} \int d\omega \left(x_+(\sqrt{s},m_1,m_2,m_3,\omega)-x_-(\sqrt{s},m_1,m_2,m_3,\omega)\right),
\ee
\noindent where
{\small 
\be
x_\pm(\sqrt{s},m_1,m_2,m_3,\omega)=\frac{\frac{\left(m_2^2-m_3^2\right) \left(\sqrt{s}-\omega \right)}{4 \sqrt{s}}\pm\frac{1}{2}\sqrt{\left(\omega ^2-m_1^2\right) \left(\omega_m\left(\sqrt{s},m_1,m_2,m_3\right)-\omega \right) \left(\frac{2 m_2m_3}{\sqrt{s}}-\omega +\omega_m\left(\sqrt{s},m_1,m_2,m_3\right)\right)}}{\frac{(m_2+m_3)^2}{2\sqrt{s}}-\omega +\omega_m\left(\sqrt{s},m_1,m_2,m_3\right)}
\ee
}
\noindent with 
\be
\omega_m(\sqrt{s},m_1,m_2,m_3)=\frac{m_1^2-(m_2+m_3)^2+s}{2 \sqrt{s}}.
\ee

Finally
\be
\Phi^{(4)}(m_X,m_\psi,m_{\pi^+},m_{\pi^-},m_{\pi^0})=\frac{1}{2\pi}\int\;ds\;\frac{1}{4\pi}\frac{p^*(m^2_X,m^2_\psi,s)}{m_{_X}}\Phi^{(3)}(\sqrt{s},m_{\pi^+},m_{\pi^-},m_{\pi^0}).
\ee
The full decay width is then
{\small \be
\begin{split}
\Gamma(X\to J/\psi\;\pi^+\pi^-\pi^0)&=\frac{1}{2s_X+1}\frac{1}{2m_{_X}}\frac{1}{6\pi}\int\;ds\;\sum_{\substack{{\rm pol}}}|\langle J/\psi\;\omega(s)|X\rangle|^2\frac{1}{4\pi}\frac{p^*(m^2_X,m^2_\psi,s)}{m_{_X}}\frac{g^2_{\omega\pi}}{(s-m_\omega^2)^2+(m_\omega\Gamma_\omega)^2}\\
&\quad\times \Phi^{(3)}(\sqrt{s},m_{\pi^+},m_{\pi^-},m_{\pi^0}).
\end{split}
\ee
}
We can relate $g^2_{\omega\pi}$ to $\Gamma(\omega\to3\pi)$
\be
g^2_{\omega\pi}=6 m_\omega\Gamma(\omega\to\pi\pi\pi)\frac{1}{\Phi^{(3)}(m_\omega,m_{\pi^+},m_{\pi^-},m_{\pi^0})},
\ee
thus giving
{\small \begin{eqnarray}
\label{gammacompletaomega}
\Gamma(X\to J/\psi\;\pi^+\pi^-\pi^0)&=&\frac{1}{2s_X+1}\frac{1}{8\pi m^2_X}\int\;ds\;\sum_{\substack{{\rm pol}}}|\langle J/\psi\;\omega(s)|X\rangle|^2p^*(m^2_X,m^2_\psi,s)\frac{1}{\pi}\frac{m_\omega\Gamma_\omega\; \mathcal{B}(\omega\to3\pi)}{(s-m_\omega^2)^2+(m_\omega\Gamma_\omega)^2}\\
&&\times \frac{\Phi^{(3)}(\sqrt{s},m_{\pi^+},m_{\pi^-},m_{\pi^0})}{\Phi^{(3)}(m_\omega,m_{\pi^+},m_{\pi^-},m_{\pi^0})}.
\end{eqnarray}
}
In the limit of narrow width for the $\omega$, Eq.~(\ref{gammacompletaomega}) is equal to the one we can obtain taking the average of $\sum_{\substack{{\rm pol}}}|\langle J/\psi\;\omega(s)|X\rangle|^2 p^*(m^2_X,m^2_\psi,s)$ over the Breit-Wigner distribution of the $\omega$ meson (using the comoving width of the $\omega$)
\be
\Gamma(X\to J/\psi\;\pi^+\pi^-\pi^0)=\frac{1}{2s_X+1}\frac{1}{8\pi m^2_X}\int\;ds\;\sum_{\substack{{\rm pol}}}|\langle J/\psi\;\omega(s)|X\rangle|^2 p^*(m^2_X,m^2_\psi,s)\frac{1}{\pi}\frac{m_\omega\Gamma_\omega\; \mathcal{B}(\omega\to\pi\pi)}{(s-m_\omega^2)^2+(m_\omega\Gamma_\omega)^2}.
\ee
In actual calculations we use $m_\omega\Gamma_\omega\to \frac{s}{m_\omega}\Gamma_\omega$.

\section{Multiplicity rules}

\subsection{$X(3872)\to D^0\bar{D}^{0*}$}
\label{app:multiplicityres}
Let us consider the decay $X(3872)\to D^0\bar{D}^{0*}$.
Since $X$ has even charge conjugation, whatever its spin is, the final state into open charm mesons needs to be
\be
|f\rangle = \frac{|D^0\bar D^{0*}\rangle+|\bar D^0 D^{0*}\rangle}{\sqrt{2}}.
\ee
The matrix element is
\be
\langle f|X\rangle = \frac{\langle D^0\bar D^{0*}|X\rangle+\langle \bar D^0 D^{0*}|X\rangle}{\sqrt{2}}.
\ee
Assuming that
\be
\langle D^0\bar D^{0*}|X\rangle=\langle \bar D^0 D^{0*}|X\rangle,
\ee
the sum over polarizations of the squared matrix element is
\be
\sum_{{\rm pol}}|\langle f|X\rangle|^2=2\sum_{{\rm pol}}|\langle D^0\bar D^{0*}|X\rangle|^2.
\ee
When we compute the cross-section for $J/\psi (\rho,\omega)\to X_J\to D^0\bar{D}^{0*}$ we actually consider the transition $J/\psi(\rho,\omega)\to f$.
The flavor wave function for the $\rho$ meson is
\be
|\rho\rangle =\frac{|u \bar u\rangle-|d \bar d\rangle }{\sqrt{2}}.
\ee
Since the neutral $D$ mesons contain only the $u$ quark ($|D^0\rangle=|c\bar u\rangle$ e $|\bar{D}^0\rangle=|\bar c u\rangle$)
only the $u\bar u$ component will contribute to the transition matrix element 
\be
\mathcal{M}=\frac{1}{\sqrt{2}}\langle f|\psi\rho\rangle.
\ee
Summing over polarization the squared matrix element one obtains
\be
\label{scattering}
\begin{split}
\sum_{{\rm pol}}|\mathcal{M}|^2&=\sum_{{\rm pol}}\frac{1}{2}|\langle f|\psi\rho\rangle|^2
=\sum_{{\rm pol}}\frac{1}{2}\times 4|\langle D^0\bar D^{0*}|\psi\rho\rangle|^2=2\sum_{{\rm pol}}|\langle D^0\bar D^{0*}|\psi\rho\rangle|^2.
\end{split}
\ee
Thus one can reabsorb the factor $2$ into the $g_{_{JDD^*}}$ coupling.

\subsection{Non resonant processes}
\label{app:multiplicitynonres}
We consider the {\it t}~-~channel processes of Eq.~(\ref{standard}).
We computed the average absorption lengths for each of the particles in the initial state, $A=\pi,\,\eta,\,\rho,\,\omega,\,\phi,\,K^{(*)}$ using the couplings defined in~\cite{Maiani:2004qj} 
\be
\label{average}
\langle \rho\sigma_{J/\psi\;A\to D\bar{D}}\rangle_{_T}=(2s_A+1)\int \frac{d^3 p_A}{(2\pi)^3}\frac{\sigma_A}{e^{E_A/\kappa_B T}-1}.
\ee
Depending on the flavor content of each meson in the initial state one can define the possible open-charm mesons configuration in the final state.
The flavor wave functions of the mesons we considered are the following~\cite{Nakamura:2010zzi}
(we neglect the $s\bar{s}$ component of the $\eta$ meson, since the contribution of the associated final state, $D^{+(*)}_s D^{-(*)}_s$, is small compared to the one coming from the $(u\bar{u}+d\bar{d})/\sqrt{2}$ component, {\it i.e.}, $D^{0(*)}\bar{D}^{0(*)}$ or  $D^{\pm(*)}D^{\mp(*)}$)
\begin{equation}
\begin{split}
&\pi^+(\rho^+) =u\bar{d},\;\pi^-(\rho^-) =\bar{u}d,\;\pi^0(\rho^0)=\frac{u\bar{u}-d\bar{d}}{\sqrt{2}}\\
&\eta\simeq\frac{u\bar{u}+d\bar{d}}{\sqrt{2}}\\
&\omega\simeq\frac{u\bar{u}+d\bar{d}}{\sqrt{2}}\\
&\phi\simeq s\bar{s}\\
&K^0=s\bar{d},\;\bar{K}^0=\bar{s}d,\;K^+=u\bar{s},\;K^-=\bar{u}s.
\end{split}
\end{equation}
The multiplicity coefficients $c^{A}_{i}$ associated to the possible final states $f_i$ for each initial particle $A$ are summarized in Table~\ref{tab:coefficients}.
Given these coefficients the total dissociation cross-section for the initial particle $A$ can be written as 
\be
\sigma_{_A}=\sum_i c^{A}_{i} \;\sigma_{AJ/\psi\to f_i}.
\ee
We can summarize all the contributions as follows
\be
\sigma_\pi=3\times \Big[\sigma(J/\psi \;\pi\to D\bar{D})+2\sigma(J/\psi \;\pi\to D\bar{D}^*)+\sigma(J/\psi \;\pi\to D^*\bar{D}^*)\Big],
\ee
\be
\sigma_{\rho+\omega}=4\times\Big[\sigma(J/\psi \;\rho\to D\bar{D})+2\sigma(J/\psi \;\rho\to D\bar{D}^*)+\sigma(J/\psi \;\rho\to D^*\bar{D}^*)\Big],
\ee
\be
\sigma_\eta=\sigma(J/\psi \;\eta\to D\bar{D})+2\sigma(J/\psi \;\eta\to D\bar{D}^*)+\sigma(J/\psi \;\eta\to D^*\bar{D}^*),
\ee
\be
\sigma_\phi=\sigma(J/\psi \;\phi\to D_s^-D_s^+)+2 \sigma(J/\psi \;\phi\to D_s^-D_s^{+*})+\sigma(J/\psi \;\phi\to D_s^{-*}D_s^{+*}),
\ee
\be
\sigma_K=4\times\Big[\sigma(J/\psi \;K\to D_s\bar{D})+\sigma(J/\psi \;K\to D_s^*\bar{D})+\sigma(J/\psi \;K\to D_s\bar{D}^*)+\sigma(J/\psi \;K\to D_s^*\bar{D}^*)\Big],
\ee
\begin{table}[ht!]
\begin{tabular}{||c|c|c|c|c|c|c|c|c|c|c|c||}
\hline
$A$ &$\bar{D}^{0(*)}D^{\pm(*)}$ &$\bar{D}^0D^{\pm *}$ &$D^{0(*)}\bar{D}^{0(*)}$ &$D^{0}\bar{D}^{0 *}$ &$D^{+(*)}D^{-(*)}$ &$D^{+}D^{- *}$ &$D_s^{+(*)}D_s^{-(*)}$ &$D_s^{+}D_s^{- *}$ &$D_s^{(*)}\bar{D}^{(*)}$ &$D_s\bar{D}^{*}$ &$D^*_s\bar D$ \\
\hline
$\pi^\pm$ &$1$ &$2$ &$0$ &$0$ &$0$ &$0$ &$0$ &$0$ &$0$ &$0$&$0$\\
\hline
$\rho^\pm$ &$1$ &$2$ &$0$ &$0$ &$0$ &$0$ &$0$ &$0$ &$0$ &$0$&$0$\\
\hline
$\pi^0$ &$0$ &$0$ &$1/2$ &$1$ &$1/2$ &$1$ &$0$ &$0$ &$0$ &$0$&$0$\\
\hline
$\rho^0,\omega$ &$0$ &$0$ &$1/2$ &$1$ &$1/2$ &$1$ &$0$ &$0$ &$0$ &$0$&$0$\\
\hline
$\eta$ &$0$ &$0$ &$1/2$ &$1$ &$1/2$ &$1$ &$0$ &$0$ &$0$ &$0$&$0$\\
\hline
$\phi$ &$0$ &$0$ &$0$ &$0$ &$0$ &$0$ &$1$ &$2$ &$0$ &$0$&$0$\\
\hline
$K^{0}$ &$0$ &$0$ &$0$ &$0$ &$0$ &$0$ &$0$ &$0$ &$1$ &$1$&$1$\\
\hline
$\bar{K}^{0}$ &$0$ &$0$ &$0$ &$0$ &$0$ &$0$ &$0$ &$0$ &$1$ &$1$&$1$\\
\hline
$K^{\pm}$ &$0$ &$0$ &$0$ &$0$ &$0$ &$0$ &$0$ &$0$ &$1$ &$1$&$1$\\
\hline
\end{tabular}
\caption{Multiplicity coefficients for the processes $J/\psi\;A\to D\bar{D}$. Notice that $D^{(*)}\bar{D}^{(*)}\equiv D\bar{D},\;D^{*}\bar{D}^{*}$.}
\label{tab:coefficients}
\end{table}

\bibliography{boiledX}         

\end{document}